\documentclass[11pt,a4paper]{article}

\usepackage{amsmath, amsthm, amssymb}
\usepackage{mathrsfs}
\usepackage{bm}
\usepackage{natbib}
\usepackage[dvips]{graphicx}

\newtheorem{thm}{Theorem}
\newtheorem{lem}{Lemma}
\newtheorem{prop}{Proposition}
\newtheorem{cor}{Corollary}
\newtheorem{definition}{Definition}
\newtheorem{remark}{Remark}
\newtheorem{example}{Example}

\newtheorem{exercise}{Exercise}
\newtheorem{report}{Report}

\newcommand{\Proof}{\noindent {\bf Proof: }\ }

\newcommand{\define}{\stackrel{\mathrm{def.}}{=}}
\newcommand{\Min}{\mathop{\mathrm{Minimize\ } }}
\newcommand{\St}{\mathrm{subject\ to\ }}
\newcommand{\emp}[1]{#1}
\newcommand{\Zero}{O}
\newcommand{\vzero}{\boldsymbol{0}}
\newcommand{\vInfty}{\boldsymbol{\infty}}
\newcommand{\Iden}{I}
\newcommand{\Coef}{A}
\newcommand{\E}{\mathrm{E}}
\newcommand{\V}{\mathrm{V}}
\newcommand{\Cov}{\mathrm{Cov}}
\newcommand{\ver}{v}
\newcommand{\subu}{u}
\newcommand{\subw}{w}
\newcommand{\obj}{y}
\newcommand{\des}{s}
\newcommand{\tre}{x}
\newcommand{\anc}{t}
\newcommand{\coef}{\alpha}
\newcommand{\sogo}{\tau}

\newcommand{\Ver}{V}
\newcommand{\subU}{U}
\newcommand{\subW}{W}
\newcommand{\Obj}{Y}
\newcommand{\Des}{S}
\newcommand{\Tre}{X}
\newcommand{\Anc}{T}
\newcommand{\Mu}{\mu}
\newcommand{\Ep}{\epsilon}

\newcommand{\Sigm}{\Sigma}
\newcommand{\pa}{\mathrm{pa}}
\newcommand{\rmv}{\mathrm{;}}
\newcommand{\builtin}[1]{#1 \rmv \pa(#1)}
\newcommand{\intervene}[1]{\tilde{#1}}
\newcommand{\objnum}{n_{\obj}}
\newcommand{\subi}{i}

\newcommand{\vobjFunc}{f}
\newcommand{\vecR}{\boldsymbol{r}}
\newcommand{\matQ}{Q}
\newcommand{\veciCoef}{\boldsymbol{\gamma}}
\newcommand{\veco}{\mathrm{vec}}
\newcommand{\Lower}{L}
\newcommand{\Upper}{U}
\newcommand{\opt}[1]{\bar{#1}}
\newcommand{\vkeisu}{\kappa}
\newcommand{\vdual}{\boldsymbol{\phi}}
\newcommand{\target}{m}
\newcommand{\mkeisu}{\lambda}

\newcommand{\SEM}{structural equation model\ }
\newcommand{\SEMstop}{structural equation model}
\newcommand{\SEMs}{structural equation models\ }
\newcommand{\SEMsstop}{structural equation models}
\newcommand{\KKT}{Karush-Kuhn-Tucker\ }
\newcommand{\iE}{\intervene{\E}}
\newcommand{\iV}{\intervene{\V}}
\newcommand{\vdualL}{\vdual_{\Lower}}
\newcommand{\vdualU}{\vdual_{\Upper}}

\newcommand{\Idenverver}{\Iden_{\ver \ver}}
\newcommand{\Idendesdes}{\Iden_{\des \des}}
\newcommand{\Identretre}{\Iden_{\tre \tre}}
\newcommand{\Idenancanc}{\Iden_{\anc \anc}}

\newcommand{\Zerotredes}{\Zero_{\tre \des}}

\newcommand{\Zeroancdes}{\Zero_{\anc \des}}
\newcommand{\Zeroanctre}{\Zero_{\anc \tre}}

\newcommand{\icoef}{\intervene{\coef}}

\newcommand{\ocoef}{\opt{\coef}}
\newcommand{\vVer}{\boldsymbol{\Ver}}
\newcommand{\vsubU}{\boldsymbol{\subU}}
\newcommand{\vsubW}{\boldsymbol{\subW}}
\newcommand{\vObj}{\boldsymbol{\Obj}}
\newcommand{\vDes}{\boldsymbol{\Des}}
\newcommand{\vTre}{\boldsymbol{\Tre}}
\newcommand{\vAnc}{\boldsymbol{\Anc}}
\newcommand{\ivTre}{\vTre}
\newcommand{\Sogo}{\boldsymbol{\sogo}}

\newcommand{\iCoef}{\intervene{\Coef}}
\newcommand{\iSogo}{\intervene{\Sogo}}
\newcommand{\avoidTre}{\vAnc\rightarrow\vDes}
\newcommand{\throughTre}{\vAnc\rightarrow\vTre\rightarrow\vDes}

\newcommand{\Coefverver}{\Coef_{\ver \ver}}
\newcommand{\Coefdesdes}{\Coef_{\des \des}}
\newcommand{\Coefdestre}{\Coef_{\des \tre}}
\newcommand{\Coefdesanc}{\Coef_{\des \anc}}

\newcommand{\Coeftretre}{\Coef_{\tre \tre}}
\newcommand{\Coeftreanc}{\Coef_{\tre \anc}}

\newcommand{\Coefancanc}{\Coef_{\anc \anc}}
\newcommand{\iCoeftreanc}{\iCoef_{\tre \anc}}
\newcommand{\Sogodesdes}{\Sogo_{\des \des}}
\newcommand{\Sogodestre}{\Sogo_{\des \tre}}
\newcommand{\Sogodesanc}{\Sogo_{\des \anc}}
\newcommand{\Sogotreanc}{\Sogo_{\tre \anc}}
\newcommand{\Sogotretre}{\Sogo_{\tre \tre}}
\newcommand{\iSogodesanc}{\iSogo_{\des \anc}}
\newcommand{\Sogoancanc}{\Sogo_{\anc \anc}}
\newcommand{\Sogoverver}{\Sogo_{\ver \ver}}
\newcommand{\vMu}{\boldsymbol{\Mu}}
\newcommand{\vMubindes}{\vMu_{\builtin{\des}}}
\newcommand{\vMubintre}{\vMu_{\builtin{\tre}}}
\newcommand{\vMubinanc}{\vMu_{\builtin{\anc}}}
\newcommand{\vMubinver}{\vMu_{\builtin{\ver}}}

\newcommand{\ivMu}{\intervene{\vMu}}

\newcommand{\ivMubintre}{\ivMu_{\builtin{\tre}}}
\newcommand{\Mubintre}{\Mu_{\builtin{\tre}}}
\newcommand{\iMu}{\intervene{\Mu}}
\newcommand{\oMu}{\opt{\Mu}}
\newcommand{\iMubintre}{\iMu_{\builtin{\tre}}}
\newcommand{\oMubintre}{\oMu_{\builtin{\tre}}}
\newcommand{\vEp}{\boldsymbol{\Ep}}
\newcommand{\vEpbindes}{\vEp_{\builtin{\des}}}
\newcommand{\vEpbintre}{\vEp_{\builtin{\tre}}}
\newcommand{\vEpbinanc}{\vEp_{\builtin{\anc}}}
\newcommand{\vEpbinver}{\vEp_{\builtin{\ver}}}

\newcommand{\ivEp}{\intervene{\vEp}}
\newcommand{\ivEpbintre}{\ivEp_{\builtin{\tre}}}

\newcommand{\Sigmbindes}{\Sigm_{\des\builtin{\des}}}
\newcommand{\Sigmbintre}{\Sigm_{\tre\builtin{\tre}}}
\newcommand{\Sigmbinanc}{\Sigm_{\anc\builtin{\anc}}}
\newcommand{\iSigm}{\intervene{\Sigma}}
\newcommand{\iSigmbintre}{\iSigm_{\tre\builtin{\tre}}}

\newcommand{\CoefLower}{\Coef_{\Lower}}
\newcommand{\CoefUpper}{\Coef_{\Upper}}
\newcommand{\vMuLower}{\vMu_{\Lower}}
\newcommand{\vMuUpper}{\vMu_{\Upper}}
\newcommand{\obji}{\obj_{\subi}}
\newcommand{\Obji}{\Obj_{\subi}}
\newcommand{\objf}{\obj_{1}}
\newcommand{\vkeisui}{\vkeisu_{\subi}}
\newcommand{\mkeisui}{\mkeisu_{\subi}}
\newcommand{\sumi}{\displaystyle{\sum_{\subi=1}^{\objnum}}\;}
\newcommand{\subihani}{\subi=1,\dots,\objnum}

\newcommand{\Sogoobjitre}{\Sogo_{\obji \tre}}

\newcommand{\Sogoobjianc}{\Sogo_{\obji \anc}}
\newcommand{\Sogoobjftre}{\Sogo_{\objf \tre}}
\newcommand{\Sogoobjfanc}{\Sogo_{\objf \anc}}
\newcommand{\iSogoobjianc}{\iSogo_{\obji \anc}}
\newcommand{\iSogoobjfanc}{\iSogo_{\objf \anc}}
\newcommand{\vecRi}{\vecR_{\subi}}
\newcommand{\matQi}{\matQ_{\subi}}
\newcommand{\vobjFunci}{\vobjFunc_{\obji}}

\newcommand{\vecCoefLower}{\boldsymbol{\coef}_{\Lower}}
\newcommand{\vecCoefUpper}{\boldsymbol{\coef}_{\Upper}}
\newcommand{\veciCoefopt}{\opt{\veciCoef}}
\newcommand{\vdualLopt}{\opt{\vdualL}}
\newcommand{\vdualUopt}{\opt{\vdualU}}
\newcommand{\coefLoweri}{{\{\vecCoefLower\}}_{\subi}}
\newcommand{\coefUpperi}{{\{\vecCoefUpper\}}_{\subi}}
\newcommand{\targeti}{\target_{\subi}}

\begin{document}
\title{Application of Mathematical Optimization Procedures to Intervention Effects in Structural Equation Models}
\author{
Kentaro Tanaka, \\
{\it Department of Industrial Engineering and Management, }\\ {\it Graduate School of Decision Science and Technology, }\\ {\it Tokyo Institute of Technology, Japan. }
\\
tanaka.k.al@m.titech.ac.jp
\vspace{2mm} \\
Atsushi Yagishita, \\
{\it Specialty Tire Process Engineering Development Department, }\\ {\it Bridgestone Corporation, Japan. }
\vspace{1mm} \\
and
\vspace{1mm} \\
Masami Miyakawa, \\
{\it Department of Industrial Engineering and Management, }\\ {\it Graduate School of Decision Science and Technology, }\\ {\it Tokyo Institute of Technology, Japan. }
}
\date{\today}
\maketitle

\begin{abstract}
For a given statistical model,
it often happens that it is necessary to
intervene the model to reduce the variances of the output variables.
In structural equation models, this can be done by changing the values of
the path coefficients by intervention.
First, we explain that the expectations and variance matrix can be decomposed into several parts in terms of the total effects.
Then, we show that an algorithm to obtain intervention method which minimizes the weighted sum of the variances
can be formulated as a convex quadratic programming.
This formulation allows us to impose boundary conditions for the intervention, so that we can find the practical solutions.
We also treat a problem to adjust the expectations on targets.
\end{abstract}


\section{Introduction}

The methods of structural equation models (SEMs) developed
by geneticists (\citet{Wright1923PassCoefficients}) and economists (\citet{Haavelmo1943SEM} and \citet{Koopmans1949SEM})
are widely used as analytical tools in a lot of fields including genetics, econometrics, social sciences and statistical quality control.
To meet the demands of the practical researchers,
thousands of studies on parameter estimation and model fitting for \SEMs have been made.

However, \SEMs are more than tools for analysis.
We can use \SEMs as tools to represent the causal relationships between the variables (\citet{Pearl2009Causality}).
If we intervene a part of the causal structure,
then the overall causal structure changes.
By using the \SEM that represents the correct causal relationships,
we can evaluate the amount of change caused by the intervention.
This means that we can compute the optimal intervention method
to minimize the variance of a variable.
\citet{KurokiMiyakawa2003CovariateSelection}, \citet{KurodaMiyakawaTanaka2006SEM} and \citet{Kuroki2008SEM}
evaluated the intervention effect for the variance of a variable
and give some methods to obtain the optimal intervention that minimizes the variance.
However, it is difficult to use their methods in practice
because they implicitly uses the impractical assumption
that the intervention can be made freely without any constraint
(e.g. we may have a bound for an intervention by changing a parameter of a structural equation because of the cost to change it).

In this paper,
we formulate the problems to obtain the optimal intervention that minimizes the variances
and to adjust the expectations as convex quadratic programmings.
This formulation enables us to easily impose boundary conditions for interventions.
To this purpose, we first introduce some ideas of decomposition of total effects in Section~\ref{sec:Decomposition_of_total_effects_and_Interventions}.
Note that the term ``decomposition of total effects''
means not only decomposition of total effects into direct and indirect effects,
but also decomposition by paths or set of variables.
We also explain that the expectations and variance matrix can be decomposed into several parts in terms of the total effects.
In Section~\ref{sec:Application-of-Mathematical-Optimization-Procedures-to-Intervention-Effects-for-Variances},
we show that the problem to obtain the optimal intervention that minimizes the variances can be formulated as a convex quadratic programming.
We also treat a problem to adjust the expectations.
Next, in Section~\ref{sec:Numerical_Experiment},
we show how the proposed algorithms given in Section~\ref{sec:Application-of-Mathematical-Optimization-Procedures-to-Intervention-Effects-for-Variances}
work by using a toy model.
Finally, we give some discussion in Section\ref{sec:Conclusion}.

\section{Decomposition of total effects and Interventions}
\label{sec:Decomposition_of_total_effects_and_Interventions}

First, in Section \ref{subsec:Structural_Equation_Models},
we briefly mention \SEMs and path diagrams, and then introduce some notations.
Next, in Section~\ref{subsec:Total_Effects_Means_and_Variances},
we introduce matrix representation of total effects and their decomposition.
The idea of decomposition of total effects is very important to consider the optimal intervention
which we will treat in Section~\ref{sec:Application-of-Mathematical-Optimization-Procedures-to-Intervention-Effects-for-Variances}.
Finally, in Section \ref{subsec:Interventions_to_Structural_Equation_Models},
we explain the interventions to the \SEMsstop.

\subsection{Structural Equation Models}
\label{subsec:Structural_Equation_Models}
The models that the relations among random variables are described in terms of linear equations are called \emp{structural equation models}.
To give some explanations about terms and notations, let us consider an example of \SEMstop.

\begin{example}
{\rm
\label{ex:five}

Assume that six random variables $\Anc_{1}, \Anc_{2}$, $\Tre_{1}$, $\Tre_{2}$, $\Des_{1}$ and $\Des_{2}$ are generated by the following linear structural equations:
\begin{eqnarray}
\Anc & = & \mu_{\builtin{\anc}} + \Ep_{\builtin{\anc}}, \nonumber \\
\Tre_{1} & = & \mu_{\builtin{\tre_{1}}} + \coef_{\tre_{1}\anc}\Anc + \Ep_{\builtin{\tre_{1}}}, \nonumber \\
\Tre_{2} & = & \mu_{\builtin{\tre_{2}}} + \coef_{\tre_{2}\anc}\Anc + \coef_{\tre_{2}\tre_{1}}\Tre_{1} + \Ep_{\builtin{\tre_{2}}}, \nonumber \\
\Des_{1} & = & \mu_{\builtin{\des_{1}}} + \coef_{\des_{1}\tre_{1}}\Tre_{1} + \Ep_{\builtin{\des_{1}}}, \nonumber \\
\Des_{2} & = & \mu_{\builtin{\des_{2}}} + \coef_{\des_{2}\anc}\Anc + \coef_{\des_{2}\tre_{2}}\Tre_{2} + \coef_{\des_{2}\des_{1}}\Des_{1} + \Ep_{\builtin{\des_{2}}}, \nonumber
\end{eqnarray}
where:
\begin{itemize}
\item $\mu_{\builtin{\anc_{1}}},\dots,\mu_{\builtin{\des_{2}}}$
are the intercepts;
\item $\coef_{\des_{1}\anc},\dots,\coef_{\des_{2}\des_{1}}$
are proportionality coefficients called \emp{path coefficients};
\item $\Ep_{\builtin{\anc}},\dots,\Ep_{\builtin{\des_{2}}}$
are the
error terms.
\end{itemize}
We will soon explain the meanings of subscripts such as ${\builtin{\tre_{2}}}$.

\bigskip
\bigskip
\begin{figure}[thbp]
\begin{center}
\includegraphics[clip,width=6.8cm]{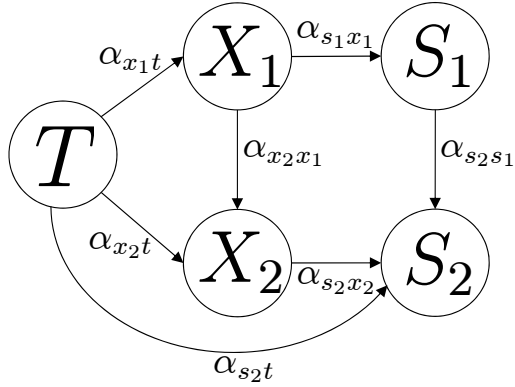}
\caption{An example of path diagram}
\label{fig:five}
\end{center}
\end{figure}

In the above equations, we presume that each left-hand side is determined by the right-hand side,
i.e. right-hand sides are causes and left-hand sides are the results.
If we represent a causal effect by an arrow with its path coefficient,
then the relations among the random variables $\Anc$, $\Tre_{1}$, $\Tre_{2}$, $\Des_{1}$ and $\Des_{2}$
can be graphically represented as Figure~\ref{fig:five}.
This graph is called the \emp{path diagram}.

The arrow from $\Anc$ to $\Tre_{2}$ means presumed direct causal effect from $\Anc$ to $\Tre_{2}$.
For this arrow, $\Anc$ is said to be \emp{parent} of $\Tre_{2}$.
Conversely, $\Tre_{2}$ is said to be \emp{child} of $\Anc$.
These are graph theoretic terms.
Here, $\Tre_{2}$ has two parents $\Anc$ and $\Tre_{1}$, and we denote them by $\pa(\tre_{2})$ as an abbreviation for parents of $\Tre_{2}$.
Furthermore, we denote by $\rmv \pa(\tre_{2})$ removing the effect of $\pa(\tre_{2})$.
Thus, $\mu_{\builtin{\tre_{2}}}$ represents the mean of $\Tre_{2}$ when the effects of the parents of $\Tre_{2}$ are removed.
We also use terms \emp{ancestor} and \emp{descendant} as graph theoretic terms.
For example, the ancestors of $\Des_{1}$ are $\Tre_{1}$ and $\Anc$,
and the descendants of $\Tre_{1}$ are $\Tre_{2}$, $\Des_{1}$  and $\Des_{2}$.
\qed
}
\end{example}

We now formulate the general \SEM in a way
so that it is easier to use for the calculations of total effects, means and variances
which we will treat in Section~\ref{subsec:Total_Effects_Means_and_Variances}.
Consider a random vector $\vVer$ the elements of which are generated by linear structural equations.
We divide the random vector $\vVer$ into three disjoint parts: $\vAnc$, $\vTre$ and $\vDes$,
so that the elements of $\vAnc$ are the ancestors of some elements of $\vTre$
and the elements of $\vDes$ are not the ancestors of some elements of $\vTre$ nor some elements of $\vTre$ themselves.
This decomposition is uniquely determined if once we choose $\vTre \subset \vVer$.

Now, a \SEM can be represented by using vectors and matrices as follows:
\begin{eqnarray}
\label{eq:linear_structural_equation_model}
\begin{pmatrix}
\vAnc \\
\vTre \\
\vDes
\end{pmatrix}
=
\begin{pmatrix}
\vMubinanc \\
\vMubintre \\
\vMubindes
\end{pmatrix}
+
\begin{pmatrix}
\Coefancanc & \Zeroanctre & \Zeroancdes \\
\Coeftreanc & \Coeftretre & \Zerotredes \\
\Coefdesanc & \Coefdestre & \Coefdesdes
\end{pmatrix}
\begin{pmatrix}
\vAnc \\
\vTre \\
\vDes
\end{pmatrix}
+
\begin{pmatrix}
\vEpbinanc \\
\vEpbintre \\
\vEpbindes
\end{pmatrix}
.
\end{eqnarray}
Here, $\vMubinanc$, $\vMubintre$ and $\vMubindes$ are the means of $\vAnc$, $\vTre$ and $\vDes$, respectively
when the effects of their parents are removed;
$\Coefancanc$, $\Coeftreanc$, $\dots$, $\Coefdesdes$ are the matrices which consist of the path coefficients;
and $\vEpbinanc$, $\vEpbintre$ and $\vEpbindes$ are the error terms.
We assume that the means of $\vEpbinanc, \vEpbintre, \vEpbindes$ are all zero values
and $\vEpbinanc, \vEpbintre, \vEpbindes$ have the variance matrices $\Sigmbinanc$, $\Sigmbintre$ and $\Sigmbindes$ respectively.
Furthermore, to avoid cycles in the structural equations,
we assume that the elements in diagonal and upper triangular portion of the coefficients matrices $\Coefancanc$, $\Coeftretre$ and $\Coefdesdes$ are all zero values.
This formulation is possible by sorting the variables by their parent-child relations whenever the structural equations do not contain cycles.
For example, the equations in Example~\ref{ex:five} can be formulated in the form of~(\ref{eq:linear_structural_equation_model})
by letting
$\vAnc = \{\Anc\}$, $\vTre = \{\Tre_{1}, \Tre_{2}\}$ and $\vDes = \{\Des_{1}, \Des_{2}\}$,
where the matrices of the path coefficients are as follows:
\begin{eqnarray}
& \Coefancanc = 0
\; , \;
\Coeftreanc = \begin{pmatrix} \coef_{\tre_{1}\anc} \\ \coef_{\tre_{2}\anc} \end{pmatrix}
\; , \;
\Coeftretre = \begin{pmatrix} 0 & 0 \\ \coef_{\tre_{2}\tre_{1}} & 0 \end{pmatrix}
\; , \;
\nonumber \\
& \Coefdesanc = \begin{pmatrix} 0 \\ \coef_{\des_{2}\anc} \end{pmatrix}
\; , \;
\Coefdestre = \begin{pmatrix} \coef_{\des_{1}\tre_{1}} & 0 \\ 0 & \coef_{\des_{2}\tre_{2}} \end{pmatrix}
\; , \;
\Coefdesdes = \begin{pmatrix} 0 & 0 \\ \coef_{\des_{2}\des_{1}} & 0 \end{pmatrix}.
\nonumber
\end{eqnarray}

\subsection{Total Effects, Means and Variances}
\label{subsec:Total_Effects_Means_and_Variances}

For a given \SEMstop,
the \emp{total effect} from a variable $\Ver_{1} \in \vVer$ to a variable $\Ver_{2} \in \vVer$ which is one of the descendants of $\Ver_{1}$
is defined as the change in $\Ver_{2}$ that is produced
when $\Ver_{1}$ is increased by $1$ and all error terms are fixed to $0$.
Therefore the total effect from $\Ver_{1}$ to $\Ver_{2}$ is equal to
the derivative of $\Ver_{2}$ with respect to $\Ver_{1}$
for the structural equations eliminating all error terms.
The \emp{direct effect} from $\Ver_{1}$ to $\Ver_{2}$
is defined as the path coefficient from $\Ver_{1}$ to $\Ver_{2}$
and it coincides with the partial derivative of $\Ver_{2}$ with respect to $\Ver_{1}$
for the structural equations eliminating all error terms.
The \emp{indirect effect} from $\Ver_{1}$ to $\Ver_{2}$
is defined as the total effect minus the direct effect.
For the precise and general definitions of the terms such as direct, indirect and total effects,
see \citet{Bollen1987TotalEffect}, \citet{Bollen1989StructuralEquations}, \citet{Sobel1990TotalEffect} and \citet{Pearl2009Causality}.
Let us consider the following example.
\begin{example}
{\rm
In Example~\ref{ex:five}, the total effect from $\Anc$ to $\Des_{2}$ is calculated as follows.

We obtain the following equations by eliminating all error terms in structural equations in Example~\ref{ex:five}.
\begin{eqnarray}
\Anc & = & \mu_{\builtin{\anc}} \nonumber \\
\Tre_{1} & = & \mu_{\builtin{\tre_{1}}} + \coef_{\tre_{1}\anc}\Anc \nonumber \\
\Tre_{2} & = & \mu_{\builtin{\tre_{2}}} + \coef_{\tre_{2}\anc}\Anc + \coef_{\tre_{2}\tre_{1}}\Tre_{1} \nonumber \\
\Des_{1} & = & \mu_{\builtin{\des_{1}}} + \coef_{\des_{1}\tre_{1}}\Tre_{1} \nonumber \\
\Des_{2} & = & \mu_{\builtin{\des_{2}}} + \coef_{\des_{2}\anc}\Anc + \coef_{\des_{2}\tre_{2}}\Tre_{2} + \coef_{\des_{2}\des_{1}}\Des_{1} \nonumber
\end{eqnarray}
From the above equations, we obtain the following relation between $\Des_{2}$ and $\Anc$
when all error terms are fixed to $0$.
\begin{eqnarray}
\Des_{2}
& = &
\mu_{\builtin{\des_{2}}} + \coef_{\des_{2}\anc}\Anc + \coef_{\des_{2}\tre_{2}}\Tre_{2} + \coef_{\des_{2}\des_{1}}\Des_{1}
\nonumber \\
& = &
\mu_{\builtin{\des_{2}}} + \coef_{\des_{2}\anc}\Anc
+ \coef_{\des_{2}\tre_{2}}(\mu_{\builtin{\tre_{2}}} + \coef_{\tre_{2}\anc}\Anc + \coef_{\tre_{2}\tre_{1}}\Tre_{1})
+ \coef_{\des_{2}\des_{1}}(\mu_{\builtin{\des_{1}}} + \coef_{\des_{1}\tre_{1}}\Tre_{1})
\nonumber \\
& = &
\mu_{\builtin{\des_{2}}} + \coef_{\des_{2}\anc}\Anc
+ \coef_{\des_{2}\tre_{2}}
\{\mu_{\builtin{\tre_{2}}} + \coef_{\tre_{2}\anc}\Anc + \coef_{\tre_{2}\tre_{1}}(\mu_{\builtin{\tre_{1}}} + \coef_{\tre_{1}\anc}\Anc)\}
\nonumber \\ & & +
\coef_{\des_{2}\des_{1}}\{\mu_{\builtin{\des_{1}}} + \coef_{\des_{1}\tre_{1}}(\mu_{\builtin{\tre_{1}}} + \coef_{\tre_{1}\anc}\Anc)\}
\nonumber \\
& = &
\mu_{\builtin{\des_{2}}} + \coef_{\des_{2}\tre_{2}}\mu_{\builtin{\tre_{2}}} + \coef_{\des_{2}\des_{1}}\mu_{\builtin{\des_{1}}}
+ (\coef_{\des_{2}\tre_{2}}\coef_{\tre_{2}\tre_{1}} + \coef_{\des_{2}\des_{1}}\coef_{\des_{1}\tre_{1}})\mu_{\builtin{\tre_{1}}}
\nonumber \\
& & +
(\coef_{\des_{2}\anc} + \coef_{\des_{2}\des_{1}}\coef_{\des_{1}\tre_{1}}\coef_{\tre_{1}\anc}
+ \coef_{\des_{2}\tre_{2}}\coef_{\tre_{2}\tre_{1}}\coef_{\tre_{1}\anc} + \coef_{\des_{2}\tre_{2}}\coef_{\tre_{2}\anc})\Anc
\nonumber
\end{eqnarray}
Therefore the total effect from $\Anc$ to $\Des_{2}$ is equal to
$\coef_{\des_{2}\anc} + \coef_{\des_{2}\des_{1}}\coef_{\des_{1}\tre_{1}}\coef_{\tre_{1}\anc} + \coef_{\des_{2}\tre_{2}}\coef_{\tre_{2}\tre_{1}}\coef_{\tre_{1}\anc} + \coef_{\des_{2}\tre_{2}}\coef_{\tre_{2}\anc}$.
The total effect can be decomposed into direct and indirect effects.
First, the direct effect  is $\coef_{\des_{2}\anc}$ which is the path coefficient of $\Anc \rightarrow \Des_{2}$.
The remainder
$\coef_{\des_{2}\des_{1}}\coef_{\des_{1}\tre_{1}}\coef_{\tre_{1}\anc}+\coef_{\des_{2}\tre_{2}}\coef_{\tre_{2}\tre_{1}}\coef_{\tre_{1}\anc}+\coef_{\des_{2}\tre_{2}}\coef_{\tre_{2}\anc}$
is the indirect effect and the terms
$\coef_{\des_{2}\des_{1}}\coef_{\des_{1}\tre_{1}}\coef_{\tre_{1}\anc}$,
$\coef_{\des_{2}\tre_{2}}\coef_{\tre_{2}\tre_{1}}\coef_{\tre_{1}\anc}$
and
$\coef_{\des_{2}\tre_{2}}\coef_{\tre_{2}\anc}$
correspond respectively to the effects of the paths
$\Anc \rightarrow \Tre_{1} \rightarrow \Des_{1} \rightarrow \Des_{2}$,
$\Anc \rightarrow \Tre_{1} \rightarrow \Tre_{2} \rightarrow \Des_{2}$
and
$\Anc \rightarrow \Tre_{2} \rightarrow \Des_{2}$
from the front.
\qed
}
\end{example}

Let us denote the total effect from $\Ver_{1} \in \vVer$ to $\Ver_{2} \in \vVer$ by $\sogo_{\ver_{2}\ver_{1}}$.
Furthermore, let us denote the matrix of the total effects from $\vsubU \subset \vVer$ to $\vsubW \subset \vVer$ by $\Sogo_{\subw\subu}$
where $\vsubU \cap \vsubW = \emptyset$ and $(i,j)$-element of $\Sogo_{\subw\subu}$ is the total effect from $\subU_{j} \in \vsubU$ to $\subW_{i} \in \vsubW$.

\begin{prop}
\label{prop:total_effect}
{\rm (\citet{Bollen1987TotalEffect}, \citet{Sobel1990TotalEffect})}
Assume that the structural equations for $\vVer$ are written in the equation
$\vVer = \vMubinver + \Coefverver\vVer + \vEpbinver$.
Furthermore, we assume that $(\Idenverver - \Coefverver)$ is invertible
where $\Idenverver$ is the identity matrix.
Then the matrix of the total effect $\Sogoverver$ is given by $\Sogoverver=(\Idenverver - \Coefverver)^{-1}\Coefverver$.
\end{prop}
Note that $(\Idenverver - \Sogoverver)^{-1}$ always exists in the model of (\ref{eq:linear_structural_equation_model}).
Intuitively, the elements of $\Coefverver$ represents the direct effects
and the elements of $\Coefverver^{2}$ represents the indirect effects through one variable.
In the same way, the elements of $\Coefverver^{n}$ can be considered as the indirect effects through $n-1$ variable.
Therefore, the total effect is equal to
$\Coefverver + \Coefverver^{2} + \Coefverver^{3} + \dots = (\Idenverver - \Coefverver)^{-1}\Coefverver$
and the above proposition holds.

In the next example,
we treat a decomposition of a total effect
and introduce some useful notations for the calculations of means and variances of $\vVer$ which we will treat later in this section.

\begin{example}
\label{ex:six}
{\rm
Assume that six random variables $\Anc_{1}$, $\Anc_{2}$, $\Tre_{1}$, $\Tre_{2}$, $\Des_{1}$ and $\Des_{2}$ are generated by the following linear structural equations:
\begin{eqnarray}
\Anc_{1} & = & \mu_{\builtin{\anc_{1}}} + \Ep_{\builtin{\anc_{1}}} \nonumber \\
\Anc_{2} & = & \mu_{\builtin{\anc_{2}}} + \coef_{\anc_{2}\anc_{1}}\Anc_{1} + \Ep_{\builtin{\anc_{2}}} \nonumber \\
\Tre_{1} & = & \mu_{\builtin{\tre_{1}}} + \coef_{\tre_{1}\anc_{1}}\Anc_{1} + \Ep_{\builtin{\tre_{1}}} \nonumber \\
\Tre_{2} & = & \mu_{\builtin{\tre_{2}}} + \coef_{\tre_{2}\anc_{1}}\Anc_{1} + \coef_{\tre_{2}\anc_{2}}\Anc_{2} + \coef_{\tre_{2}\tre_{1}}\Tre_{1} + \Ep_{\builtin{\tre_{2}}} \nonumber \\
\Des_{1} & = & \mu_{\builtin{\des_{1}}} + \coef_{\des_{1}\anc_{1}}\Anc_{1} + \coef_{\des_{1}\tre_{1}}\Tre_{1} + \Ep_{\builtin{\des_{1}}} \nonumber \\
\Des_{2} & = & \mu_{\builtin{\des_{2}}} + \coef_{\des_{2}\anc_{1}}\Anc_{1} + \coef_{\des_{2}\anc_{2}}\Anc_{2} + \coef_{\des_{2}\tre_{1}}\Tre_{1} + \coef_{\des_{2}\tre_{2}}\Tre_{2} + \coef_{\des_{2}\des_{1}}\Des_{1} + \Ep_{\builtin{\des_{2}}} \nonumber
\end{eqnarray}
The path diagram of the above linear structural equations is given in Figure~\ref{fig:six}.

\bigskip
\begin{figure}[thbp]
\begin{center}
\includegraphics[clip,width=6.8cm]{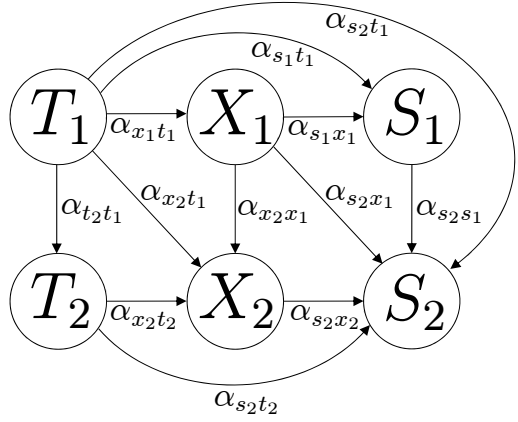}
\caption{An example of path diagram with six variables}
\label{fig:six}
\end{center}
\end{figure}

The above equations can be formulated in the form of~(\ref{eq:linear_structural_equation_model})
by letting
$\vAnc = \{\Anc_{1}, \Anc_{2}\}$, $\vTre = \{\Tre_{1}, \Tre_{2}\}$ and $\vDes = \{\Des_{1}, \Des_{2}\}$,
where the matrices of the path coefficients are as follows:
\begin{eqnarray}
\Coefancanc = \begin{pmatrix} 0 & 0 \\ \coef_{\anc_{2}\anc_{1}} & 0\end{pmatrix}
\; , \;
\Coeftreanc = \begin{pmatrix} \coef_{\tre_{1}\anc_{1}} & 0 \\ \coef_{\tre_{2}\anc_{1}} & \coef_{\tre_{2}\anc_{2}} \end{pmatrix}
\; , \;
\Coeftretre = \begin{pmatrix} 0 & 0 \\ \coef_{\tre_{2}\tre_{1}} & 0\end{pmatrix}
\nonumber \\
\Coefdesanc = \begin{pmatrix} \coef_{\des_{1}\anc_{1}} & 0 \\ \coef_{\des_{2}\anc_{1}} & \coef_{\des_{2}\anc_{2}} \end{pmatrix}
\; , \;
\Coefdestre = \begin{pmatrix} \coef_{\des_{1}\tre_{1}} & 0 \\ \coef_{\des_{2}\tre_{1}} & \coef_{\des_{2}\tre_{2}} \end{pmatrix}
\; , \;
\Coefdesdes = \begin{pmatrix} 0 & 0 \\ \coef_{\des_{2}\des_{1}} & 0 \end{pmatrix}.
\nonumber
\end{eqnarray}

In this model,
the total effect from $\Anc_{1}$ to $\Des_{2}$ is calculated as follows:
\begin{eqnarray}
\lefteqn{
\sogo_{\des_{2}\anc_{1}}
=
\coef_{\des_{2}\anc_{1}}
+
\coef_{\des_{2}\des_{1}}\coef_{\des_{1}\anc_{1}}
+
\coef_{\des_{2}\anc_{2}}\coef_{\anc_{2}\anc_{1}}
+
\coef_{\des_{2}\tre_{1}}\coef_{\tre_{1}\anc_{1}}
+
{{\coef_{\des_{2}\tre_{2}}}\coef_{\tre_{2}\anc_{1}}}}
& & \nonumber \\
& & \qquad \qquad \qquad \qquad \qquad
+
{\coef_{\des_{2}\tre_{2}}}{\coef_{\tre_{2}\tre_{1}}}{\coef_{\tre_{1}\anc_{1}}}
+
{\coef_{\des_{2}\des_{1}}}{\coef_{\des_{1}\tre_{1}}}{\coef_{\tre_{1}\anc_{1}}}
+
{\coef_{\des_{2}\tre_{2}}}{\coef_{\tre_{2}\anc_{2}}}{\coef_{\anc_{2}\anc_{1}}}.
\nonumber
\end{eqnarray}
Furthermore, the total effect from $\Anc_{1}$ to $\Des_{2}$ is decomposed into the following eight paths:
\begin{eqnarray}
& \Anc_{1} \xrightarrow{\coef_{\des_{2}\anc_{1}}} \Des_{2}, \nonumber \\
& \Anc_{1} \xrightarrow{\coef_{\des_{1}\anc_{1}}} \Des_{1} \xrightarrow{\coef_{\des_{2}\des_{1}}} \Des_{2}, \nonumber \\
& \Anc_{1} \xrightarrow{\coef_{\anc_{2}\anc_{1}}} \Anc_{2} \xrightarrow{\coef_{\des_{2}\Anc_{2}}} \Des_{2}, \nonumber \\
& \Anc_{1} \xrightarrow{\coef_{\tre_{1}\anc_{1}}} \Tre_{1} \xrightarrow{\coef_{\des_{2}\tre_{1}}} \Des_{2}, \nonumber \\
& \Anc_{1} \xrightarrow{\coef_{\tre_{2}\anc_{1}}} \Tre_{2} \xrightarrow{\coef_{\des_{2}\tre_{2}}} \Des_{2}, \label{eq:paths_sixmodel}\\
& \Anc_{1} \xrightarrow{\coef_{\tre_{1}\anc_{1}}} \Tre_{1} \xrightarrow{\coef_{\tre_{2}\tre_{1}}} \Tre_{2} \xrightarrow{\coef_{\des_{2}\tre_{2}}} \Des_{2}, \nonumber \\
& \Anc_{1} \xrightarrow{\coef_{\tre_{1}\anc_{1}}} \Tre_{1} \xrightarrow{\coef_{\des_{1}\tre_{1}}} \Des_{1} \xrightarrow{\coef_{\des_{2}\des_{1}}} \Des_{2}, \nonumber \\
& \Anc_{1} \xrightarrow{\coef_{\anc_{2}\anc_{1}}} \Anc_{2} \xrightarrow{\coef_{\tre_{2}\anc_{2}}} \Tre_{2} \xrightarrow{\coef_{\des_{2}\tre_{2}}} \Des_{2}. \nonumber
\end{eqnarray}
In the above paths, only the first path $\Anc_{1} \xrightarrow{\coef_{\des_{2}\anc_{1}}} \Des_{2}$ represents
the direct effect with the value of $\coef_{\des_{2}\anc_{1}}$
and the other paths represent indirect effects with the values of
$
\coef_{\des_{2}\des_{1}}\coef_{\des_{1}\anc_{1}}
$, $
\coef_{\des_{2}\anc_{2}}\coef_{\anc_{2}\anc_{1}}
$, $
\coef_{\des_{2}\tre_{1}}\coef_{\tre_{1}\anc_{1}}
$, $
\coef_{\des_{2}\tre_{1}}\coef_{\tre_{1}\anc_{1}}
$, $
\coef_{\des_{2}\tre_{2}}\coef_{\tre_{2}\anc_{1}}
$, $
{\coef_{\des_{2}\tre_{2}}}{\coef_{\tre_{2}\tre_{1}}}{\coef_{\tre_{1}\anc_{1}}}
$ and $
{\coef_{\des_{2}\des_{1}}}{\coef_{\des_{1}\tre_{1}}}{\coef_{\tre_{1}\anc_{1}}}
$, $
{\coef_{\des_{2}\tre_{2}}}{\coef_{\tre_{2}\anc_{2}}}{\coef_{\anc_{2}\anc_{1}}}
$
respectively.

\bigskip
\begin{figure}[ttbp]
\vspace{0.65cm}
\begin{center}
\includegraphics[clip,width=6.8cm]{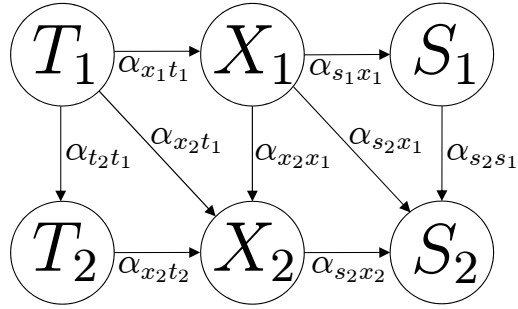}
\end{center}
\vspace{0.15cm}
\caption{A path diagram when the direct paths from $\Anc$ to $\Des$ are removed.}
\label{fig:sixThroughX}
\end{figure}
\begin{figure}[thbp]
\begin{center}
\includegraphics[clip,width=6.8cm]{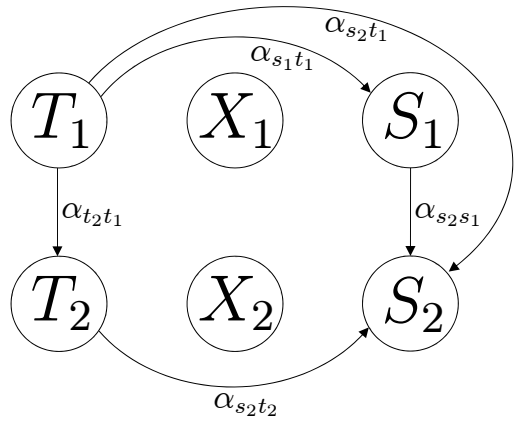}
\end{center}
\vspace{-0.5cm}
\caption{A path diagram when the paths through $\tre$ are removed.}
\label{fig:sixAvoidX}
\end{figure}

Now,
we decompose the total effect from $\Anc_{1}$ to $\Des_{2}$ into the following two parts.

\begin{enumerate}
\item Let us denote by $\sogo_{\des_{2}\anc_{1}}(\throughTre)$
the total effect from $\Anc_{1}$ to $\Des_{2}$ through $\vTre$.
Because the last five paths in~(\ref{eq:paths_sixmodel})
go through $\Tre_{1}$ and $\Tre_{2}$,
we obtain
\begin{eqnarray}
\sogo_{\des_{2}\anc_{1}}(\throughTre)
& = &
\coef_{\des_{2}\tre_{1}}\coef_{\tre_{1}\anc_{1}}
+
\coef_{\des_{2}\tre_{1}}\coef_{\tre_{1}\anc_{1}}
+
\coef_{\des_{2}\tre_{2}}\coef_{\tre_{2}\anc_{1}}
\nonumber \\
& &
+
{\coef_{\des_{2}\tre_{2}}}{\coef_{\tre_{2}\tre_{1}}}{\coef_{\tre_{1}\anc_{1}}}
+
{\coef_{\des_{2}\des_{1}}}{\coef_{\des_{1}\tre_{1}}}{\coef_{\tre_{1}\anc_{1}}}
+
{\coef_{\des_{2}\tre_{2}}}{\coef_{\tre_{2}\anc_{2}}}{\coef_{\anc_{2}\anc_{1}}}.
\nonumber
\end{eqnarray}
This is equal to the total effect from $\Anc_{1}$ to $\Des_{2}$ in the model of Figure~\ref{fig:sixThroughX}.
\item Let us denote by $\sogo_{\des_{2}\anc_{1}}(\avoidTre)$
the total effect from $\Anc_{1}$ to $\Des_{2}$ when the effect of  $\vTre$ are removed.
From the above decomposition, the first three paths in~(\ref{eq:paths_sixmodel})
do not go through $\Tre_{1}$ and $\Tre_{2}$.
Therefore, we obtain
\begin{equation}
\sogo_{\des_{2}\anc_{1}}(\avoidTre)
=
\coef_{\des_{2}\anc_{1}}
+
\coef_{\des_{2}\des_{1}}\coef_{\des_{1}\anc_{1}}
+
\coef_{\des_{2}\anc_{2}}\coef_{\anc_{2}\anc_{1}}.
\nonumber
\end{equation}
This is equal to the total effect from $\Anc_{1}$ to $\Des_{2}$ in the model of Figure~\ref{fig:sixAvoidX}.
\end{enumerate}

Next, let us consider the following two matrices
\begin{eqnarray}
\Sogodesanc(\throughTre) & \define & (\Idendesdes - \Coefdesdes)^{-1}\Coefdestre(\Identretre - \Coeftretre)^{-1}\Coeftreanc(\Idenancanc - \Coefancanc)^{-1},
\nonumber \\
\Sogodesanc(\avoidTre) & \define & (\Idendesdes - \Coefdesdes)^{-1}\Coefdesanc(\Idenancanc - \Coefancanc)^{-1},
\nonumber
\end{eqnarray}
where $\Idenancanc, \Identretre$ and $\Idendesdes$ are the identity matrices.
Then we obtain
\begin{eqnarray}
\lefteqn{\Sogodesanc(\throughTre)
} & &
\nonumber \\
& = &
\begin{pmatrix} 1 & 0 \\ -\coef_{\des_{2}\des_{1}} & 1 \end{pmatrix}^{-1}
\begin{pmatrix} \coef_{\des_{1}\tre_{1}} & 0 \\ \coef_{\des_{2}\tre_{1}} & \coef_{\des_{2}\tre_{2}} \end{pmatrix}
\begin{pmatrix} 1 & 0 \\ -\coef_{\tre_{2}\tre_{1}} & 1 \end{pmatrix}^{-1}
\begin{pmatrix} \coef_{\tre_{1}\anc_{1}} & 0 \\ \coef_{\tre_{2}\anc_{1}} & \coef_{\tre_{2}\anc_{2}} \end{pmatrix}
\begin{pmatrix} 1 & 0 \\ -\coef_{\anc_{2}\anc_{1}} & 1 \end{pmatrix}^{-1}
\nonumber \\
& = &
\begin{pmatrix} 1 & 0 \\ \coef_{\des_{2}\des_{1}} & 1 \end{pmatrix}
\begin{pmatrix} \coef_{\des_{1}\tre_{1}} & 0 \\ \coef_{\des_{2}\tre_{1}} & \coef_{\des_{2}\tre_{2}} \end{pmatrix}
\begin{pmatrix} 1 & 0 \\ \coef_{\tre_{2}\tre_{1}} & 1 \end{pmatrix}
\begin{pmatrix} \coef_{\tre_{1}\anc_{1}} & 0 \\ \coef_{\tre_{2}\anc_{1}} & \coef_{\tre_{2}\anc_{2}} \end{pmatrix}
\begin{pmatrix} 1 & 0 \\ \coef_{\anc_{2}\anc_{1}} & 1 \end{pmatrix}
\nonumber \\
& = &
\begin{pmatrix}
\coef_{\des_{1}\tre_{1}}\coef_{\tre_{1}\anc_{1}} &  0 \\
\left\{
\begin{array}{c}
\coef_{\des_{2}\tre_{1}}\coef_{\tre_{1}\anc_{1}}
+
{\coef_{\des_{2}\tre_{1}}}{\coef_{\tre_{1}\anc_{1}}}
+
{\coef_{\des_{2}\tre_{2}}}{\coef_{\tre_{2}\anc_{1}}}
\\
+
{\coef_{\des_{2}\tre_{2}}}{\coef_{\tre_{2}\tre_{1}}}{\coef_{\tre_{1}\anc_{1}}}
+
{\coef_{\des_{2}\des_{1}}}{\coef_{\des_{1}\tre_{1}}}{\coef_{\tre_{1}\anc_{1}}}
+
{\coef_{\des_{2}\tre_{2}}}{\coef_{\tre_{2}\anc_{2}}}{\coef_{\anc_{2}\anc_{1}}}
\end{array}
\right\}
& \coef_{\des_{2}\tre_{2}}\coef_{\tre_{2}\anc_{2}}
\end{pmatrix},
\nonumber
\end{eqnarray}
and
\begin{eqnarray}
\Sogodesanc(\avoidTre)
& = &
\begin{pmatrix} 1 & 0 \\ -\coef_{\des_{2}\des_{1}} & 1 \end{pmatrix}^{-1}
\begin{pmatrix} \coef_{\des_{1}\anc_{1}} & 0 \\ \coef_{\des_{2}\anc_{1}} & \coef_{\des_{2}\anc_{2}} \end{pmatrix}
\begin{pmatrix} 1 & 0 \\ -\coef_{\anc_{2}\anc_{1}} & 1 \end{pmatrix}^{-1}
\nonumber \\
& = &
\begin{pmatrix} 1 & 0 \\ \coef_{\des_{2}\des_{1}} & 1 \end{pmatrix}
\begin{pmatrix} \coef_{\des_{1}\anc_{1}} & 0 \\ \coef_{\des_{2}\anc_{1}} & \coef_{\des_{2}\anc_{2}} \end{pmatrix}
\begin{pmatrix} 1 & 0 \\ \coef_{\anc_{2}\anc_{1}} & 1 \end{pmatrix}
\nonumber \\
& = &
\begin{pmatrix}
\coef_{\des_{1}\anc_{1}} &  0 \\
\coef_{\des_{2}\anc_{1}} + \coef_{\des_{2}\des_{1}}\coef_{\des_{1}\anc_{1}} + \coef_{\des_{2}\anc_{2}}\coef_{\anc_{2}\anc_{1}}
& \coef_{\des_{2}\anc_{2}}
\end{pmatrix}.
\nonumber
\end{eqnarray}
Note that the $(2,1)$-elements of $\Sogodesanc(\throughTre)$ and $\Sogodesanc(\avoidTre)$,
which corresponds to $(\Des_{2}, \Anc_{1}$),
are equivalent to $\sogo_{\des_{2}\anc_{1}}(\throughTre)$ and $\sogo_{\des_{2}\anc_{1}}(\avoidTre)$.
This equivalence can be justified as Theorem~\ref{thm:decomposition_of_total_effect}.
}\qed
\end{example}

As in Example~\ref{ex:six}, we define the following two matrices for the model of (\ref{eq:linear_structural_equation_model}):
\begin{eqnarray}
\Sogodesanc(\throughTre) & \define & (\Idendesdes - \Coefdesdes)^{-1}\Coefdestre(\Identretre - \Coeftretre)^{-1}\Coeftreanc(\Idenancanc - \Coefancanc)^{-1},
\label{eq:Sogodesanc_throughTre} \\
\Sogodesanc(\avoidTre) & \define & (\Idendesdes - \Coefdesdes)^{-1}\Coefdesanc(\Idenancanc - \Coefancanc)^{-1},
\label{eq:Sogodesanc_avoidTre}
\end{eqnarray}
where $\Idenancanc, \Identretre$ and $\Idendesdes$ are the identity matrices.
The next lemma can be shown by direct calculation.
\begin{lem}
Let $B$ be a square matrix which can be represented as follows:
\begin{equation}
B = \left(
\begin{array}{c|c|c}
B_{11} & \Zero & \Zero \\ \hline
B_{21} & B_{22} & \Zero \\ \hline
B_{31} & B_{32} & B_{33}
\end{array}
\right),
\nonumber
\end{equation}
where $B_{11},B_{22},B_{33}$ are square matrices.
If $B_{11}, B_{22}, B_{33}$ are non-singular matrices,
then the following equation holds for the inverse matrix of $B$.
\begin{equation}
\left(
\begin{array}{c|c|c}
B_{11} & \Zero & \Zero \\ \hline
B_{21} & B_{22} & \Zero \\ \hline
B_{31} & B_{32} & B_{33}
\end{array}
\right)^{-1}
=
\left(
\begin{array}{c|c|c}
B_{11}^{-1} & \Zero & \Zero \\ \hline
-B_{22}^{-1}B_{21}B_{11}^{-1} & B_{22}^{-1} & \Zero \\ \hline
B_{33}^{-1}B_{32}B_{22}^{-1}B_{21}B_{11}^{-1}-B_{33}^{-1}B_{31}B_{11}^{-1} & -B_{33}^{-1}B_{32}B_{22}^{-1} & B_{33}^{-1} \\
\end{array}
\right)
\nonumber
\end{equation}
\label{lem:lower_triangle}
\qed
\end{lem}

In the next theorem, we obtain the matrix representations of total effects
from $\vAnc$ to $\vTre$, from $\vTre$ to $\vDes$ and from $\vAnc$ to $\vDes$,
and justify the decomposition of the total effect which is treated in Example~\ref{ex:six}.
\begin{thm}
\label{thm:decomposition_of_total_effect}
\begin{eqnarray}
\Sogotreanc
& = &
[(\Iden - \Coef)^{-1} \Coef]_{\tre\anc}
=
(\Identretre - \Coeftretre)^{-1} \Coeftreanc (\Idenancanc - \Coefancanc)^{-1},
\label{eq:thm_total_effect_from_anc_to_tre} \\
\Sogodestre
& = &
[(\Iden - \Coef)^{-1} \Coef]_{\des\tre}
=
(\Idendesdes - \Coefdesdes)^{-1} \Coefdestre (\Identretre - \Coeftretre)^{-1},
\label{eq:thm_total_effect_from_tre_to_des} \\
\Sogodesanc
& = &
[(\Iden - \Coef)^{-1} \Coef]_{\des\anc}
\nonumber \\
& = &
(\Idendesdes - \Coefdesdes)^{-1} \Coefdestre (\Identretre - \Coeftretre)^{-1} \Coeftreanc (\Idenancanc - \Coefancanc)^{-1}
+
(\Idendesdes - \Coefdesdes)^{-1} \Coefdesanc (\Idenancanc - \Coefancanc)^{-1}
\label{eq:thm_total_effect_from_anc_to_des} \\
& = &
\Sogodesanc(\throughTre) + \Sogodesanc(\avoidTre)
\label{eq:thm_decomposition_of_total_effect}
\end{eqnarray}
where $[(\Iden - \Coef)^{-1} \Coef]_{\subu\subw}$ for $\vsubU, \vsubW \in \vVer$ is the submatrix of $(\Iden - \Coef)^{-1} \Coef$
corresponding to the rows of $\vsubU$ and the columns of $\vsubW$.
\end{thm}
\Proof
By letting
$
B_{11} = \Idenancanc-\Coefancanc
\; , \;
B_{21} = -\Coeftreanc
\; , \;
B_{22} = \Identretre-\Coeftretre
\; , \;
B_{31} = -\Coefdesanc
\; , \;
B_{32} = -\Coefdestre
\; , \;
B_{33} = \Idendesdes-\Coefdesdes
$
in Lemma~\ref{lem:lower_triangle},
we obtain
\begin{equation}
\left(
I - \Coef
\right)^{-1}
=
\left(
\begin{array}{c|c|c}
(\Idenancanc-\Coefancanc)^{-1} & \Zero & \Zero \\ \hline
(\Identretre-\Coeftretre)^{-1}\Coeftreanc(\Idenancanc-\Coefancanc)^{-1} & (\Identretre-\Coeftretre)^{-1} & \Zero \\ \hline
{\Coefdesanc^{\ast}}
& (\Idendesdes-\Coefdesdes)^{-1}\Coefdestre(\Identretre-\Coeftretre)^{-1} & (\Idendesdes-\Coefdesdes)^{-1}
\end{array}
\right),
\nonumber
\end{equation}
where
\begin{equation}
{\Coefdesanc^{\ast}} =
(\Idendesdes-\Coefdesdes)^{-1}\Coefdestre(\Identretre-\Coeftretre)^{-1}\Coeftreanc(\Idenancanc-\Coefancanc)^{-1}+(\Idendesdes-\Coefdesdes)^{-1}\Coefdesanc(\Idenancanc-\Coefancanc)^{-1}.
\nonumber
\end{equation}
By using the definitions of $\Sogodesanc(\throughTre)$ and $\Sogodesanc(\avoidTre)$ in (\ref{eq:Sogodesanc_throughTre}) and (\ref{eq:Sogodesanc_avoidTre}),
and the identity $(\Iden-C)^{-1}C+I=(\Iden-C)^{-1}$ for non-singular matrix $C$, we obtain
{\small
\begin{equation}
\left(
I - \Coef
\right)^{-1}
\Coef
=
\left(
\begin{array}{c|c|c}
(\Idenancanc-\Coefancanc)^{-1}\Coefancanc & \Zero & \Zero \\ \hline
(\Identretre-\Coeftretre)^{-1}\Coeftreanc(\Idenancanc-\Coefancanc)^{-1} & (\Identretre-\Coeftretre)^{-1}\Coeftretre & \Zero \\ \hline
\Sogodesanc(\throughTre) + \Sogodesanc(\avoidTre) & (\Idendesdes-\Coefdesdes)^{-1}\Coefdestre(\Identretre-\Coeftretre)^{-1} & (\Idendesdes-\Coefdesdes)^{-1}\Coefdesdes
\end{array}
\right).
\nonumber
\end{equation}
}
Therefore, we obtain
the matrix representations (\ref{eq:thm_total_effect_from_anc_to_tre}),~(\ref{eq:thm_total_effect_from_tre_to_des}) and~(\ref{eq:thm_total_effect_from_anc_to_des}),
and the decomposition $\Sogodesanc = \Sogodesanc(\throughTre) + \Sogodesanc(\avoidTre)$.
\qed

Note that
\begin{equation}
\Sogoancanc = (\Idenancanc-\Coefancanc)^{-1}\Coefancanc
\, , \,
\Sogotretre = (\Identretre-\Coeftretre)^{-1}\Coeftretre
\, , \,
\Sogodesdes = (\Idendesdes-\Coefdesdes)^{-1}\Coefdesdes
\label{eq:total_effect_within_subset}
\end{equation}
are also obtained from the proof of Theorem~\ref{thm:decomposition_of_total_effect},
and they are also obtained from Proposition~\ref{prop:total_effect}.
Furthermore, note that, for example,  the matrix of total effects $\Sogotreanc$ needs both
the premultiplication of $(\Identretre-\Coeftretre)^{-1}$ and the postmultiplication of $(\Idenancanc - \Coefancanc)^{-1}$.
This is a thing that is different from the result of Proposition~\ref{prop:total_effect}.

Next, we calculate the means of $\vAnc$, $\vTre$ and $\vDes$.
From \SEM (\ref{eq:linear_structural_equation_model}),
we obtain the following equations:
\begin{eqnarray}
(\Idenancanc - \Coefancanc)\vAnc & = & \vMubinanc + \vEpbinanc, \nonumber \\
(\Identretre - \Coeftretre)\vTre & = & \vMubintre \Coeftreanc\vAnc + \vEpbintre, \nonumber \\
(\Idendesdes - \Coefdesdes)\vDes & = & \vMubindes + \Coefdestre\vTre + \Coefdesanc\vAnc + \vEpbindes. \nonumber
\end{eqnarray}
By multiplying both sides of the above three equations
by inverse of $(\Idenancanc - \Coefancanc), (\Identretre - \Coeftretre)$ and $(\Idendesdes - \Coefdesdes)$ respectively,
we obtain the following equations:
\begin{eqnarray}
\vAnc & = & (\Idenancanc - \Coefancanc)^{-1}\vMubinanc + (\Idenancanc - \Coefancanc)^{-1}\vEpbinanc, \label{eq:transformed_sem_anc} \\
\vTre & = & (\Identretre - \Coeftretre)^{-1}\vMubintre + (\Identretre - \Coeftretre)^{-1}\Coeftreanc\vAnc + (\Identretre - \Coeftretre)^{-1}\vEpbintre, \label{eq:transformed_sem_tre} \\
\vDes & = & (\Idendesdes - \Coefdesdes)^{-1}\vMubindes  + (\Idendesdes - \Coefdesdes)^{-1}\Coefdesanc\vAnc + (\Idendesdes - \Coefdesdes)^{-1}\Coefdestre\vTre + (\Idendesdes - \Coefdesdes)^{-1}\vEpbindes.
\nonumber \\ \label{eq:transformed_sem_des}
\end{eqnarray}
By taking the means of both sides of (\ref{eq:transformed_sem_anc}),~(\ref{eq:transformed_sem_tre}) and~(\ref{eq:transformed_sem_des}),
we can compute the means of $\vAnc$, $\vTre$ and $\vDes$, and obtain the following proposition.
\begin{prop}
\label{prop:expectation_of_descendants}
\begin{eqnarray}
\E[\vAnc]
& = &
(\Sogoancanc + \Idenancanc)\vMubinanc,
\label{eq:prop_mean_anc}
\\
\E[\vTre]
& = &
\Sogotreanc\vMubinanc + (\Sogotretre + \Identretre)\vMubintre,
\label{eq:prop_mean_tre}
\\
\E[\vDes]
& = & \Sogodesanc\vMubinanc + \Sogodestre\vMubintre + (\Sogodesdes + \Idendesdes)\vMubindes
\label{eq:prop_mean_des}
\end{eqnarray}
\end{prop}
\Proof
By taking the means of both sides of (\ref{eq:transformed_sem_anc})
and using~(\ref{eq:total_effect_within_subset}),
we obtain~(\ref{eq:prop_mean_anc}) as follows:
\begin{equation}
\E[\vAnc]
= (\Idenancanc - \Coefancanc)^{-1}\vMubinanc
= (\Idenancanc - \Coefancanc)^{-1}\{\Coefancanc + (\Idenancanc - \Coefancanc)\}\vMubinanc
= (\Sogoancanc + \Idenancanc)\vMubinanc.
\nonumber
\end{equation}

Next, by substituting (\ref{eq:transformed_sem_anc}) into (\ref{eq:transformed_sem_tre}) and taking the means,
we obtain (\ref{eq:prop_mean_tre}) as follows:
\begin{eqnarray}
\E[\vTre]
& = &
(\Identretre - \Coeftretre)^{-1}\Coeftreanc\E[\vAnc] + (\Identretre - \Coeftretre)^{-1}\vMubintre
\nonumber \\
& = &
(\Identretre - \Coeftretre)^{-1}\Coeftreanc(\Idenancanc - \Coefancanc)^{-1}\vMubinanc + (\Identretre - \Coeftretre)^{-1}\vMubintre
\nonumber \\
& = &
\Sogotreanc\vMubinanc + (\Sogotretre + \Identretre)\vMubintre,
\nonumber
\end{eqnarray}
where we are using (\ref{eq:thm_total_effect_from_anc_to_tre}) and (\ref{eq:total_effect_within_subset}) in the third equality.

Finally, by substituting (\ref{eq:transformed_sem_anc}) and (\ref{eq:transformed_sem_tre}) into (\ref{eq:transformed_sem_des}) and taking the means,
we obtain (\ref{eq:prop_mean_des}) as follows:
\begin{eqnarray}
\E[\vDes]
& = &
(\Idendesdes - \Coefdesdes)^{-1}\Coefdestre
\{(\Identretre - \Coeftretre)^{-1}\Coeftreanc(\Idenancanc - \Coefancanc)^{-1}\vMubinanc + (\Identretre - \Coeftretre)^{-1}\vMubintre\}
\nonumber \\
& & +
(\Idendesdes - \Coefdesdes)^{-1}\Coefdesanc(\Idenancanc - \Coefancanc)^{-1}\vMubinanc
+ (\Idendesdes - \Coefdesdes)^{-1}\vMubindes
\nonumber \\
& = &
\{\Sogodesanc(\throughTre) + \Sogodesanc(\avoidTre)\}\vMubinanc
+ \Sogodestre\vMubintre
+ (\Sogodesdes + \Idendesdes)\vMubindes,
\nonumber \\
& = &
\Sogodesanc\vMubinanc
+ \Sogodestre\vMubintre
+ (\Sogodesdes + \Idendesdes)\vMubindes,
\nonumber
\end{eqnarray}
where we are using
(\ref{eq:Sogodesanc_throughTre}),~(\ref{eq:Sogodesanc_avoidTre}),~(\ref{eq:thm_total_effect_from_tre_to_des}) and (\ref{eq:total_effect_within_subset})
in the second equality
and using (\ref{eq:thm_decomposition_of_total_effect})
in the third equality.
\qed

The above proposition says that the means can be decomposed by means of the total effects.
Almost the same things can be said about the variance matrix of $\vAnc$, $\vTre$ and $\vDes$.
\begin{prop}
\label{prop:variance_of_descendants}
Assume that $\Cov[\vAnc, \vEpbintre] = \Cov[\vAnc, \vEpbindes] = \Cov[\vTre, \vEpbindes] = \Zero$.
\begin{eqnarray}
\V[\vAnc]
& = &
(\Sogoancanc + \Idenancanc)\Sigmbinanc(\Sogoancanc + \Idenancanc)^{T},
\label{eq:prop_var_anc}
\\
\V[\vTre]
& = &
\Sogotreanc\Sigmbinanc\Sogotreanc^{T}
+ (\Sogotretre + \Identretre)\Sigmbintre(\Sogotretre + \Identretre)^{T},
\label{eq:prop_var_tre}
\\
\V[\vDes]
& = &
\Sogodesanc\Sigmbinanc\Sogodesanc^{T}
+ \Sogodestre\Sigmbintre\Sogodestre^{T}
+ (\Sogodesdes + \Idendesdes)\Sigmbindes(\Sogodesdes + \Idendesdes)^{T}
\label{eq:prop_var_des}
\end{eqnarray}
\end{prop}
\Proof

From (\ref{eq:transformed_sem_anc}),
we obtain (\ref{eq:prop_var_anc}) as follows:
\begin{eqnarray}
\V[\vAnc]
& = &
(\Idenancanc - \Coefancanc)^{-1}\V[\vEpbinanc](\Idenancanc - \Coefancanc)^{-T}
\label{eq:proof_var_anc_1}\\
& = &
(\Idenancanc - \Coefancanc)^{-1}\{\Coefancanc + (\Idenancanc - \Coefancanc)\}\V[\vEpbinanc]\{\Coefancanc + (\Idenancanc - \Coefancanc)\}^{T}(\Idenancanc - \Coefancanc)^{-T}
\nonumber \\
& = & (\Sogoancanc + \Idenancanc)\Sigmbinanc(\Sogoancanc + \Idenancanc)^{T}
\nonumber
\end{eqnarray}
where we are using (\ref{eq:total_effect_within_subset}) and $\V[\vEpbinanc] = \Sigmbinanc$ in the third equality.

Next, from
(\ref{eq:transformed_sem_anc}), (\ref{eq:transformed_sem_tre}), (\ref{eq:proof_var_anc_1})
and the assumption $\Cov[\vAnc, \vEpbintre] = \Zero$,
we obtain (\ref{eq:prop_var_tre}) as follows:
\begin{eqnarray}
\V[\vTre]
& = &
(\Identretre - \Coeftretre)^{-1}\Coeftreanc\V[\vAnc]\Coeftreanc^{T}(\Identretre - \Coeftretre)^{-T}
+
(\Identretre - \Coeftretre)^{-1}\V[\vEpbintre](\Identretre - \Coeftretre)^{-T}
\nonumber \\
& = &
\{(\Identretre - \Coeftretre)^{-1}\Coeftreanc(\Idenancanc - \Coefancanc)^{-1}\}\V[\vEpbinanc]\{(\Identretre - \Coeftretre)^{-1}\Coeftreanc(\Idenancanc - \Coefancanc)^{-1}\}^{T}
\nonumber \\
& & +
(\Identretre - \Coeftretre)^{-1}\V[\vEpbintre](\Identretre - \Coeftretre)^{-T}
\label{eq:proof_var_tre_1} \\
& = &
\Sogotreanc\Sigmbinanc\Sogotreanc^{T}
+ (\Sogotretre + \Identretre)\Sigmbintre(\Sogotretre + \Identretre)^{T},
\nonumber
\end{eqnarray}
where we are using (\ref{eq:thm_total_effect_from_anc_to_tre}), (\ref{eq:total_effect_within_subset}) and $\V[\vEpbintre] = \Sigmbintre$ in the third equality.

Finally, from the assumption $\Cov[\vAnc, \vEpbindes] = \Cov[\vTre, \vEpbindes] = \Zero$,
we have
\begin{eqnarray}
\lefteqn{\V[\vDes]
=
(\Idendesdes - \Coefdesdes)^{-1}\Coefdesanc\V[\vAnc]\Coefdesanc^{T}(\Idendesdes - \Coefdesdes)^{-T}
} & & \nonumber \\
& & +
(\Idendesdes - \Coefdesdes)^{-1}\Coefdesanc\Cov[\vAnc, \vTre]\Coefdestre^{T}(\Idendesdes - \Coefdesdes)^{-T}
+
(\Idendesdes - \Coefdesdes)^{-1}\Coefdestre\Cov[\vTre, \vAnc]\Coefdesanc^{T}(\Idendesdes - \Coefdesdes)^{-T}
\nonumber \\
& & +
(\Idendesdes - \Coefdesdes)^{-1}\Coefdestre\V[\vTre]\Coefdestre^{T}(\Idendesdes - \Coefdesdes)^{-T}
\nonumber \\
& & +
(\Idendesdes - \Coefdesdes)^{-1}\V[\vEpbindes](\Idendesdes - \Coefdesdes)^{-T}.
\label{eq:proof_var_des_1}
\end{eqnarray}
Now, from (\ref{eq:transformed_sem_anc}) and (\ref{eq:transformed_sem_tre}), $\Cov[\vAnc, \vTre]^{T}=\Cov[\vTre, \vAnc]$ can be calculated as follows:
\begin{eqnarray}
\Cov[\vTre, \vAnc]
& = & (\Identretre - \Coeftretre)^{-1}\Coeftreanc\V[\vAnc]
= (\Identretre - \Coeftretre)^{-1}\Coeftreanc(\Idenancanc - \Coefancanc)\V[\vEpbinanc]
\label{eq:proof_cov_tre_anc} \\
& = & \Sogotreanc\Sigmbinanc
\nonumber
\end{eqnarray}
Therefore, by substituting (\ref{eq:proof_var_anc_1}), (\ref{eq:proof_var_tre_1}) and (\ref{eq:proof_cov_tre_anc}) into (\ref{eq:proof_var_des_1}),
we obtain (\ref{eq:prop_var_des}) as follows:
{\footnotesize
\begin{eqnarray}
\lefteqn{\V[\vDes]
=
\{(\Idendesdes - \Coefdesdes)^{-1}\Coefdesanc(\Idenancanc - \Coefancanc)^{-1}\}\V[\vEpbinanc]\{(\Idendesdes - \Coefdesdes)^{-1}\Coefdesanc(\Idenancanc - \Coefancanc)^{-1}\}^{T}
} & & \nonumber \\
& & +
\{(\Idendesdes - \Coefdesdes)^{-1}\Coefdestre(\Identretre - \Coeftretre)^{-1}\Coeftreanc(\Idenancanc - \Coefancanc)\}\V[\vEpbinanc]\{(\Idendesdes - \Coefdesdes)^{-1}\Coefdesanc(\Idenancanc - \Coefancanc)^{-1}\}^{T}
\nonumber \\
& & +
\{(\Idendesdes - \Coefdesdes)^{-1}\Coefdesanc(\Idenancanc - \Coefancanc)^{-1}\}\V[\vEpbinanc]\{(\Idendesdes - \Coefdesdes)^{-1}\Coefdestre(\Identretre - \Coeftretre)^{-1}\Coeftreanc(\Idenancanc - \Coefancanc)\}^{T}
\nonumber \\
& & +
\{(\Idendesdes - \Coefdesdes)^{-1}\Coefdestre(\Identretre - \Coeftretre)^{-1}\Coeftreanc(\Idenancanc - \Coefancanc)^{-1}\}\V[\vEpbinanc]\{(\Idendesdes - \Coefdesdes)^{-1}\Coefdestre(\Identretre - \Coeftretre)^{-1}\Coeftreanc(\Idenancanc - \Coefancanc)^{-1}\}^{T}
\nonumber \\
& & +
\{(\Idendesdes - \Coefdesdes)^{-1}\Coefdestre(\Identretre - \Coeftretre)^{-1}\}\V[\vEpbintre]\{(\Idendesdes - \Coefdesdes)^{-1}\Coefdestre(\Identretre - \Coeftretre)^{-1}\}^{T}
\nonumber \\
& & +
(\Idendesdes - \Coefdesdes)^{-1}\V[\vEpbindes](\Idendesdes - \Coefdesdes)^{-T}
\nonumber \\
& = &
\Sogodesanc(\avoidTre)\Sigmbinanc\Sogodesanc(\avoidTre)^{T}
\nonumber \\
& & +
\Sogodestre(\throughTre)\Sigmbinanc\Sogodesanc(\avoidTre)^{T}
+
\Sogodesanc(\avoidTre)\Sigmbinanc\Sogodestre(\throughTre)^{T}
\nonumber \\
& & +
\Sogodesanc(\throughTre)\Sigmbinanc\Sogodesanc(\throughTre)^{T}
\nonumber \\
& & +
\Sogodestre\Sigmbintre\Sogodestre^{T}
\nonumber \\
& & +
(\Sogodesdes + \Idendesdes)\Sigmbindes(\Sogodesdes + \Idendesdes)^{T}
\nonumber  \\
& = &
\Sogodesanc\Sigmbinanc\Sogodesanc^{T}
+
\Sogodestre\Sigmbintre\Sogodestre^{T}
+
(\Sogodesdes + \Idendesdes)\Sigmbindes(\Sogodesdes + \Idendesdes)^{T},
\nonumber
\end{eqnarray}
}
where we are using (\ref{eq:Sogodesanc_throughTre}), (\ref{eq:Sogodesanc_avoidTre}), (\ref{eq:thm_total_effect_from_anc_to_tre}), (\ref{eq:thm_total_effect_from_tre_to_des}), (\ref{eq:total_effect_within_subset}),
$\V[\vEpbinanc]=\Sigmbinanc$, $\V[\vEpbintre]=\Sigmbintre$ and $\V[\vEpbindes] = \Sigmbindes$
in the second equality,
and (\ref{eq:thm_decomposition_of_total_effect}) in the third equality.
\qed

In the following, we only consider the case where the assumption of Proposition~\ref{prop:variance_of_descendants} holds, i.e.
$\Cov[\vAnc, \vEpbintre] = \Cov[\vAnc, \vEpbindes] = \Cov[\vTre, \vEpbindes] = \Zero$.

\subsection{Interventions to Structural Equation Models}
\label{subsec:Interventions_to_Structural_Equation_Models}

An \emp{intervention} to a \SEM means changing structure of the \SEMstop.
Throughout this paper,
we consider only intervention to the structures between $\vAnc$ and $\vTre$ in the model of (\ref{eq:linear_structural_equation_model}),
(for more general case of intervention, see \citet{Pearl2009Causality}).
In this case, only the elements of $\vTre$ are directly affected by the intervention and are called \emp{treatment variables}.
Of course, the elements of $\vDes$ are also affected indirectly by the intervention.
The elements of $\vAnc$ are called \emp{covariates} and the elements of $\vDes$ are called \emp{output variables}.
The effects caused by the intervention are called \emp{intervention effects}.
For example, the changes on the means of the output variables $\vDes$ after the intervention are intervention effects.

Assume that $\vMubintre$, $\Coeftreanc$ and $\vEpbintre$ in (\ref{eq:linear_structural_equation_model})
are changed into $\ivMubintre$, $\iCoeftreanc$ and $\ivEpbintre$, respectively, by the intervention,
where $\ivEpbintre$ is the column vector of error terms that their means are all zero values and the variance matrix is $\iSigmbintre$.
Furthermore, we assume that the assumption of Proposition~\ref{prop:variance_of_descendants} again holds after the intervention, i.e.
$\Cov[\vAnc, \ivEpbintre] = \Cov[\vAnc, \vEpbindes] = \Cov[\vTre, \vEpbindes] = \Zero$.
Then the structural equation for $\vTre$ is changed from
\begin{eqnarray}
\vTre = \vMubintre + \Coeftretre\vTre + \Coeftreanc\vAnc + \vEpbintre
\nonumber
\end{eqnarray}
to
\begin{eqnarray}
\label{eq:definition_of_intervention_for_arrows}
\ivTre = \ivMubintre + \Coeftretre\ivTre + \iCoeftreanc\vAnc + \ivEpbintre.
\end{eqnarray}

Let us define the following matrices.
\begin{eqnarray}
\iSogodesanc(\throughTre) & \define & (\Idendesdes - \Coefdesdes)^{-1}\Coefdestre(\Identretre - \Coeftretre)^{-1}\iCoeftreanc(\Idenancanc - \Coefancanc)^{-1},
\label{eq:def_iSogodesanc_throughTre} \\
\iSogodesanc & \define & \iSogodesanc(\throughTre) + \Sogodesanc(\avoidTre)
\label{eq:def_iSogodesanc}
\end{eqnarray}
The elements of $\iSogodesanc$ are the total effects from $\vAnc$ to $\vDes$ after the intervention of (\ref{eq:definition_of_intervention_for_arrows}).
Note that $\Sogodesanc(\avoidTre)$ does not change after the intervention of (\ref{eq:definition_of_intervention_for_arrows}).

Let us denote by $\iE[\vDes]$ and $\iV[\vDes]$ the means and the variances of $\vDes$
after the intervention of (\ref{eq:definition_of_intervention_for_arrows}).
Then the following corollary holds immediately from Proposition~\ref{prop:expectation_of_descendants} and~\ref{prop:variance_of_descendants}.
\begin{cor}
After the intervention of (\ref{eq:definition_of_intervention_for_arrows}),
the mean vector of $\vDes$ is given by
\label{cor:expectation_and_variance_of_descendants_with_intervention}
\begin{eqnarray}
\iE[\vDes]
& = &
\iSogodesanc\vMubinanc
+ \Sogodestre\ivMubintre + (\Sogodesdes + \Idendesdes)\vMubinanc.
\nonumber
\end{eqnarray}
Furthermore, assume that $\Cov[\vAnc, \ivEpbintre] = \Cov[\vAnc, \vEpbindes] = \Cov[\vTre, \vEpbindes] = \Zero$,
then the variance matrix of $\vDes$ after the intervention of (\ref{eq:definition_of_intervention_for_arrows}) is given by
\begin{eqnarray}
\iV[\vDes]
& = &
\iSogodesanc\Sigmbinanc\iSogodesanc^{T}
+ \Sogodestre\iSigmbintre\Sogodestre^{T}
+ (\Sogodesdes + \Idendesdes)\Sigmbindes(\Sogodesdes + \Idendesdes)^{T}.
\label{eq:variance_of_descendants_with_intervention}
\end{eqnarray}
\end{cor}

In the following sections,
we treat only the intervention by which the error terms of $\vTre$ do not change i.e. $\iSigmbintre = \Sigmbintre$.

\section{Application of Mathematical Optimization Procedures to Intervention Effects}
\label{sec:Application-of-Mathematical-Optimization-Procedures-to-Intervention-Effects-for-Variances}

In Section~\ref{subsec:Application-of-Mathematical-Optimization-Procedures-to-Intervention-Effects-for-Variances},
we first consider intervention to the path coefficients $\iSogodesanc$
to reduce the variances of the output variables.
Next, in Section~\ref{subsec:Application-of-Mathematical-Optimization-Procedures-to-Intervention-Effects-for-Means},
we treat intervention to the means $\ivMubintre$ to adjust the mean values of output variables.

\subsection{Application of Mathematical Optimization Procedures to Intervention Effects for Variances}
\label{subsec:Application-of-Mathematical-Optimization-Procedures-to-Intervention-Effects-for-Variances}

For a given \SEMstop,
it often happens that it is necessary to intervene the model to reduce the variances of the output variables.
In \SEMsstop, this can be done by changing the values of
the path coefficients $\iSogodesanc$ by intervention.
In this section, we show that an algorithm to obtain the intervention method which minimizes
the weighted sum of the variances can be formulated as a convex quadratic programming.
This formulation allows us to impose the boundary conditions
for the intervention, so that we can find the practical solutions.

Let us denote by $\vObj$
elements of interest in $\vDes$
, by $\objnum$ the dimension of $\Obji$,
and by $\Obji$ the $\subi$-th element of $\vObj$.
The variance of $\Obji$, which we denote by $\iV [\Obji]$, is the diagonal element of $\iV[\vDes]$ in relation to $\Obji$.
Then the minimization of the weighted sum of the variances of $\vObj$, under constraint
that the elements of $\iCoeftreanc$ have upper and lower bounds
can be formulated as follows:
\begin{eqnarray}
\Min_{\iCoeftreanc} & \sumi \vkeisui \iV [\Obji]
\label{eq:obj_func_var_org}
\\
\St & \CoefLower \leq \iCoeftreanc \leq \CoefUpper.
\label{eq:const_var_org}
\end{eqnarray}
where
$\vkeisu_{1},\dots,\vkeisu_{\objnum}$ are the weights,
and $\CoefLower$ and $\CoefUpper$ are the matrices,
the elements of which are the lower and upper bounds for $\iCoeftreanc$.
We assume that these values are determined appropriately in advance.

Now we formulate the above problem as a convex quadratic programming.

At first, we neglect the terms
$\Sogodestre\iSigmbintre\Sogodestre^{T}$ and $(\Sogodesdes + \Idendesdes)\Sigmbindes(\Sogodesdes + \Idendesdes)^{T}$
in the variance matrix of (\ref{eq:variance_of_descendants_with_intervention}),
because they are not changed by changing $\iCoeftreanc$.
Let us define the following functions:
\begin{eqnarray}
\vobjFunc_{\obji}(\iCoeftreanc)
& \define &
\iSogoobjianc\Sigmbinanc\iSogoobjianc^{T}
\nonumber \\
& = &
\{\iSogoobjianc(\throughTre) + \Sogoobjianc(\avoidTre)\}\Sigmbinanc\{\iSogoobjianc(\throughTre) + \Sogoobjianc(\avoidTre)\}^{T},
\nonumber \\
& & \qquad \qquad \qquad \qquad \qquad \qquad \qquad \qquad \qquad \qquad \qquad \qquad \qquad (\subihani),
\label{eq:partial_obj_func}
\end{eqnarray}
where $\iSogoobjianc(\throughTre)$ and $\Sogoobjianc(\avoidTre)$ are the row vectors
of $\iSogodesanc(\throughTre)$ and $\Sogodesanc(\avoidTre)$ in relation to $\Obji$.
Then the minimization of the objective function in (\ref{eq:obj_func_var_org}) is equivalent to
\begin{equation}
\Min_{\iCoeftreanc} \sumi \vkeisui \vobjFunc_{\obji}(\iCoeftreanc).
\nonumber
\end{equation}
Remember that the definition of $\iSogodesanc(\throughTre)$ is
$\iSogodesanc(\throughTre) = (\Idendesdes - \Coefdesdes)^{-1}\Coefdestre(\Identretre - \Coeftretre)^{-1}\iCoeftreanc(\Idenancanc - \Coefancanc)^{-1}$
in~(\ref{eq:def_iSogodesanc_throughTre}).
By using $\veco$ operator, Kronecker product $\otimes$ and~(\ref{eq:vec_and_otimes}) (see Appendix~\ref{app:Kronecker_product_and_Vec_Operator}),
the column vector $\iSogoobjianc$ in~(\ref{eq:partial_obj_func}) can be formulated as follows:
\begin{eqnarray}
\iSogoobjianc^{T}
& = & \{\iSogoobjianc(\throughTre) + \Sogoobjianc(\avoidTre)\}^{T}
\nonumber \\
& = & \left\{[(\Idendesdes - \Coefdesdes)^{-1}]_{\obji\des}\Coefdestre(\Identretre - \Coeftretre)^{-1}\iCoeftreanc(\Idenancanc - \Coefancanc)^{-1}\right\}^{T}
+ \Sogoobjianc(\avoidTre)^{T}
\nonumber \\
& = &
\veco\left(
[(\Idendesdes - \Coefdesdes)^{-1}]_{\obji\des}\Coefdestre(\Identretre - \Coeftretre)^{-1}\iCoeftreanc(\Idenancanc - \Coefancanc)^{-1}
\right)
+ \Sogoobjianc(\avoidTre)^{T}
\nonumber \\
& = &
\left[
\left\{(\Idenancanc - \Coefancanc)^{-1}\right\}^{T}
\otimes
\left\{[(\Idendesdes - \Coefdesdes)^{-1}]_{\obji\des}\Coefdestre(\Identretre - \Coeftretre)^{-1}\right\}
\right]
\veco(\iCoeftreanc)
+ \Sogoobjianc(\avoidTre)^{T},
\nonumber \\
& = &
\left[
\left\{(\Idenancanc - \Coefancanc)^{-1}\right\}^{T}
\otimes
\Sogoobjitre
\right]
\veco(\iCoeftreanc)
+ \Sogoobjianc(\avoidTre)^{T},
\label{eq:vec_in_partial_obj_func_1}
\end{eqnarray}
where $[(\Idendesdes - \Coefdesdes)^{-1}]_{\obji\des}$ is the row vector of $(\Idendesdes - \Coefdesdes)^{-1}$ in relation to $\Obji$,
and $\Sogoobjitre$ is the row vector of $\Sogodestre$ in relation to $\Obji$
(see (\ref{eq:thm_total_effect_from_tre_to_des}) of Theorem~\ref{thm:decomposition_of_total_effect}).
Let us define the following matrices and column vectors:
\begin{eqnarray*}
\matQi & \define &
\left\{(\Idenancanc - \Coefancanc)^{-1}\right\}^{T}
\otimes
\Sogoobjitre
\nonumber \\
& = &
\left\{(\Idenancanc - \Coefancanc)^{-1}\right\}^{T}
\otimes
\left\{[(\Idendesdes - \Coefdesdes)^{-1}]_{\obji\des}\Coefdestre(\Identretre - \Coeftretre)^{-1}\right\}
\; , \; (\subihani), \\
\veciCoef & \define & \veco(\iCoeftreanc), \\
\vecRi & \define & \Sogoobjianc(\avoidTre)^{T}
= \left\{[(\Idendesdes - \Coefdesdes)^{-1}]_{\obji\des}\Coefdesanc(\Idenancanc - \Coefancanc)^{-1}\right\}^{T}
\;, \; (\subihani).
\end{eqnarray*}
By using these definitions and~(\ref{eq:vec_in_partial_obj_func_1}),
the column vector $\iSogoobjianc$ in~(\ref{eq:partial_obj_func}) can be
represented as follows:
\begin{eqnarray}
\iSogoobjianc^{T}
=
\matQi\veciCoef + \vecRi.
\label{eq:vec_in_partial_obj_func_2}
\end{eqnarray}
Hence, we obtain
\begin{equation}
\vobjFunci(\iCoeftreanc)
= (\matQi\veciCoef + \vecRi)^{T} \Sigmbinanc (\matQi\veciCoef + \vecRi)
= \veciCoef^{T}\left(\matQi^{T}\Sigmbinanc\matQi\right)\veciCoef
+ \left(2\vecRi^{T}\Sigmbinanc\matQi\right)\veciCoef
+ \vecRi^{T}\Sigmbinanc\vecRi ,
\nonumber
\end{equation}
and
\begin{eqnarray}
\sumi \vkeisui \vobjFunci(\iCoeftreanc)
= \veciCoef^{T} \left(\sumi \vkeisui\matQi^{T}\Sigmbinanc\matQi\right) \veciCoef
+ \left(\sumi 2\vkeisui\vecRi^{T}\Sigmbinanc\matQi\right)\veciCoef
+ \sumi \vkeisui\vecRi^{T}\Sigmbinanc\vecRi
\nonumber
\end{eqnarray}
Note that the third term in the right-hand side of the above equation is constant with respect to $\veciCoef$
and negligible in the minimization problem of (\ref{eq:obj_func_var_org}).
Therefore, the minimization problem of (\ref{eq:obj_func_var_org})
under the constraint of (\ref{eq:const_var_org})
can be represented as the following convex quadratic programming:
\begin{eqnarray}
\Min_{\veciCoef} &
\veciCoef^{T} \left(\sumi \vkeisui\matQi^{T}\Sigmbinanc\matQi\right) \veciCoef
+ \left(\sumi 2\vkeisui\vecRi^{T}\Sigmbinanc\matQi\right)\veciCoef
\nonumber \\
\label{eq:QP_programming_for_variances} \\
\St & \vecCoefLower \leq \veciCoef \leq \vecCoefUpper.
\nonumber
\end{eqnarray}
where
$\vecCoefLower \define \veco(\CoefLower)$ and $\vecCoefUpper \define \veco(\CoefUpper)$.

The Karush-Kuhn-Tucker (KKT) conditions of the problem of~(\ref{eq:QP_programming_for_variances})
are given as follows:
\begin{eqnarray}
& \left(\sumi 2\vkeisui\matQi^{T}\Sigmbinanc\matQi\right) \veciCoef
+ \left(\sumi 2\vkeisui\vecRi^{T}\Sigmbinanc\matQi\right)^{T} -\vdualL+\vdualU = \vzero,
\nonumber \\
& \vdualL \geq \vzero
\quad , \quad
\vdualU \geq \vzero,
\nonumber \\
& -\veciCoef + \vecCoefLower \leq \vzero
\quad , \quad
\veciCoef - \vecCoefUpper \leq \vzero,
\nonumber \\
& \vdualL^{T}(-\veciCoef + \vecCoefLower) = 0
\quad , \quad
\vdualU^{T}(\veciCoef - \vecCoefUpper) = 0,
\nonumber
\end{eqnarray}
where the elements of $\vdualL$ and  $\vdualU$ are Lagrange multipliers (for more detail see \citet{RockafellarConvexAnalysis}).
Assume that the constraints in (\ref{eq:QP_programming_for_variances}) satisfy Slater's constraint qualification,
i.e. $\vecCoefLower < \vecCoefUpper$ holds.
Then $\veciCoefopt$ is optimal
if and only if there exist $\vdualLopt$ and $\vdualUopt$
which satisfy the above \KKT conditions for $\veciCoefopt$.
Notice that even if $\coefLoweri = \coefUpperi$ holds for some $i$'s,
the constraints in (\ref{eq:QP_programming_for_variances}) satisfy Slater's constraint qualification
by considering the inequality constraints as equality constraints.
\begin{example}
\label{ex:KKT_condition_for_simple_case}
{\rm
Let us consider a case where $\vObj=\{\Obj_{1}\}$,
$\Sogoobjftre
= [(\Idendesdes - \Coefdesdes)^{-1}]_{\objf\des}\Coefdestre(\Identretre - \Coeftretre)^{-1}
\neq \vzero$,
$\Sigmbinanc$ is regular,
and constraint is not imposed on $\iCoeftreanc$, i.e.
\begin{equation}
\vdualL = \vzero \;,\; \vdualU = \vzero \;,\; \vecCoefLower \rightarrow -\vInfty \;,\; \vecCoefUpper \rightarrow \vInfty.
\nonumber
\end{equation}
Then the \KKT conditions in this case are given as follows:
\begin{eqnarray}
\lefteqn{
\left(2\matQ_{1}^{T}\Sigmbinanc\matQ_{1}\right) \veciCoef
+ \left(2\vecR_{1}^{T}\Sigmbinanc\matQ_{1}\right)^{T}
= \vzero
} & &
\nonumber \\
& \Leftrightarrow &
\matQ_{1}^{T}\Sigmbinanc(\matQ_{1}\veciCoef + \vecR_{1}) = \vzero
\nonumber \\
& \Leftrightarrow &
\matQ_{1}\veciCoef + \vecR_{1} = \vzero
\qquad (\matQ_{1}\text{ is regular from the assumption }\Sogoobjftre \neq \vzero)
\nonumber \\
& \Leftrightarrow &
\iSogoobjianc = \iSogoobjianc(\throughTre) + \Sogoobjianc(\avoidTre) = \vzero.
\qquad (\text{See (\ref{eq:vec_in_partial_obj_func_2}) and (\ref{eq:def_iSogodesanc}).})
\nonumber
\end{eqnarray}
Remember that the first term $\iSogoobjfanc(\throughTre)$ in the last equation means the total effect
from $\vAnc$ to $\Obj_{1}$ through $\vTre$ after intervention
and the second term $\Sogoobjfanc(\avoidTre)$ means the total effect
from $\vAnc$ to $\Obj_{1}$ which does not go through $\vTre$.
Therefore, if the total effect from $\vAnc$ to $\Obj_{1}$ through $\vTre$ after intervention
offsets the total effect from $\vAnc$ to $\Obj_{1}$ which does not go through $\vTre$,
then the \KKT conditions hold and the variance of $\Obj_{1}$ is minimized.
}
\qed
\end{example}

\subsection{Application of Mathematical Optimization Procedures to Intervention Effects for Means}
\label{subsec:Application-of-Mathematical-Optimization-Procedures-to-Intervention-Effects-for-Means}

We consider the intervention to the means $\ivMubintre$.
From proposition~\ref{prop:expectation_of_descendants}, we obtain the mean of $\Obji$ as follows:
\begin{eqnarray}
\E[\Obji]
& = & \Sogoobjianc\vMubinanc + \Sogoobjitre\ivMubintre + [(\Sogodesdes + \Idendesdes)]_{\objf\des}\vMubindes,
\nonumber
\end{eqnarray}
where $[(\Sogodesdes + \Idendesdes)]_{\objf\des}$ is the row vector of $(\Sogodesdes + \Idendesdes)$ in relation to $\Obji$.

Suppose that we want to adjust the mean of $\Obji$ to a standard $\targeti$ by intervention which changes $\ivMubintre$.
Then the minimization of weighted squared sum of the deviations $(\E[\Obj_{1}]-\target_{1}),\dots,(\E[\Obj_{\objnum}]-\target_{\objnum})$,
under constraint that the elements of $\ivMubintre$ have upper and lower bounds
can be formulated as follows:
\begin{eqnarray}
\Min_{\ivMubintre} & \sumi \mkeisui (\iE[\Obji] - \targeti)^{2}
\label{eq:obj_func_mu_org}
\\
\St & \vMuLower \leq \ivMubintre \leq \vMuUpper.
\label{eq:const_mu_org}
\end{eqnarray}
where
$\mkeisu_{1},\dots,\mkeisu_{\objnum}$ are the weights,
and  $\vMuLower$ and $\vMuUpper$ are the matrices,
the elements of which are the lower and upper bounds for $\ivMubintre$.
We assume that these values are determined appropriately in advance.

From Proposition~\ref{prop:expectation_of_descendants}, we obtain
\begin{eqnarray}
(\iE[\Obji] - \targeti)^{2}
& = &
\{(\Sogoobjianc\vMubinanc + \Sogoobjitre\ivMubintre + [(\Sogodesdes + \Idendesdes)]_{\objf\des}\vMubindes) - \targeti\}^{2}
\nonumber \\
& = &
\ivMubintre^{T} (\Sogoobjitre^{T}\Sogoobjitre) \ivMubintre\
\nonumber \\
& & +
\left[2\{(\Sogoobjianc\vMubinanc +  [(\Sogodesdes + \Idendesdes)]_{\objf\des}\vMubindes - \targeti\}\Sogoobjitre\right]\ivMubintre
\nonumber \\
& & +
\{\Sogoobjianc\vMubinanc +  [(\Sogodesdes + \Idendesdes)]_{\objf\des}\vMubindes - \targeti\}^{2}
\nonumber
\end{eqnarray}
Note that the third term of the last equation does not depend on $\ivMubintre$ and
only the first and second terms are needed for the minimization in (\ref{eq:obj_func_mu_org}).
Therefore, the minimization problem of (\ref{eq:obj_func_mu_org})
under the constraint of (\ref{eq:const_mu_org})
can be represented as the following convex quadratic programming:
\begin{eqnarray}
\Min_{\ivMubintre} &
\ivMubintre^{T} (\Sogoobjitre^{T}\Sogoobjitre) \ivMubintre\
+
\left[2\{(\Sogoobjianc\vMubinanc +  [(\Sogodesdes + \Idendesdes)]_{\objf\des}\vMubindes - \targeti\}\Sogoobjitre\right]\ivMubintre
\nonumber \\
\nonumber \\
\St & \vMuLower \leq \ivMubintre \leq \vMuUpper.
\nonumber
\end{eqnarray}

\section{Numerical Experiment}
\label{sec:Numerical_Experiment}

\bigskip
\begin{figure}[thbp]
\begin{center}
\includegraphics[clip,width=12cm]{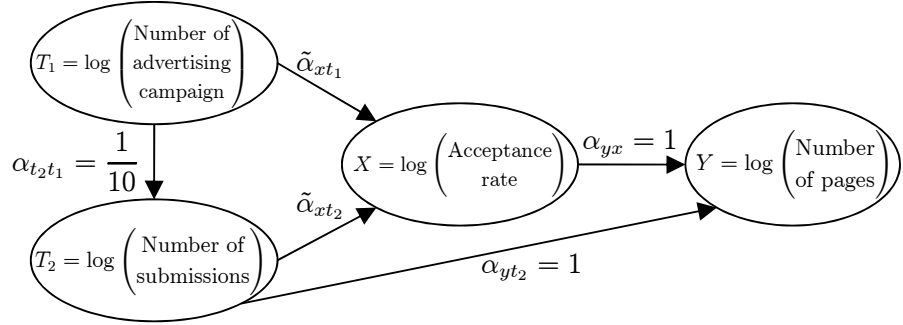}
\caption{The path diagram of the structural equation model of (\ref{eq:sem_for_numerical_experiment}).}
\label{fig:numerical_experiment}
\end{center}
\end{figure}

To illustrate how the two algorithms
in Section~\ref{sec:Application-of-Mathematical-Optimization-Procedures-to-Intervention-Effects-for-Variances}
work, we consider the following toy model.
The model used in this numerical experiment is just a toy.
It may contain some inappropriate formulations and should not be taken seriously.

Suppose that an editor of a journal
which is published once a year
wanted to stabilize the number of pages of the journal.
The editor observed the following four variables:
\begin{itemize}
\item $\Anc_{1}$ - the random variable of the logarithm of the number of advertising campaign for the journal;
\item $\Anc_{2}$ - the random variable of the logarithm of the number of submissions to the journal;
\item $X$ - the random variable of the logarithm of the acceptance rate of the journal;
\item $Y$ - the random variable of the logarithm of the number of pages of the journal.
\end{itemize}
The editor can control the borderline whether or not to accept a manuscript graded by some referees.
However, the acceptance rate is random variable
because the grades of the manuscripts submitted to the journal are determined by the reviewers.
Furthermore, the advertising campaign is not the editor's job and the editor can not control.
To these variables, the editor constructed a simplified \SEM
which is represented as the path diagram in Figure~\ref{fig:numerical_experiment} and the following equations:
\begin{eqnarray}
\Anc_{1} & = & \mu_{\builtin{\anc_{1}}} + \Ep_{\builtin{\anc_{1}}}, \nonumber \\
\Anc_{2} & = & \mu_{\builtin{\anc_{2}}} + \coef_{\anc_{2}\anc_{1}}\Anc_{1} + \Ep_{\builtin{\anc_{2}}}, \nonumber \\
\Tre & = & \mu_{\builtin{\tre}} + \coef_{\tre\anc_{1}}\Anc_{1} + \coef_{\tre\anc_{2}}\Anc_{2} + \Ep_{\builtin{\tre}},
\label{eq:sem_for_numerical_experiment} \\
\Obj & = & \mu_{\builtin{\obj}} + \coef_{\obj\anc_{2}}\Anc_{2} + \coef_{\obj\tre}\Tre + \Ep_{\builtin{\obj}}, \nonumber
\end{eqnarray}
where
\begin{itemize}
\item {\small $\vMu_{\builtin{\anc}} = \begin{pmatrix} \log{10} \\ \log{100} \end{pmatrix}
\quad
\left(
\begin{array}{l}
\text{Average number of advertising campaign is } 10,
\\ \text{and that of submissions is } 100 \text{ where the effect of the parent is removed}
\end{array}\right)$},
\item $\Mu_{\builtin{\tre}} = \log{\frac{3}{10}} \quad (\text{Average of acceptance rate is } \frac{3}{10} \text{ when the effect of } \vAnc \text{ are removed})$,
\item $\Mu_{\builtin{\obj}} = \log{10}, \quad (\text{Average number of pages for each manuscript is } 10 \text{})$,
\item $\Ep_{\builtin{\anc_{1}}}$, $\Ep_{\builtin{\anc_{2}}}$, $\Ep_{\builtin{\tre}}$ and $\Ep_{\builtin{\obj}} \sim N\left(0, \left(\frac{1}{\sqrt{10}}\right)^{2}\right)$,
\item $\coef_{\anc_{2}\anc_{1}} = \frac{1}{10}$ and $\coef_{\obj\anc_{2}} = \coef_{\obj\tre} = 1$.
\end{itemize}
The last equation in~(\ref{eq:sem_for_numerical_experiment})
means that the number of pages of the journal is
approximately equal to
$\{${Average number of pages for each manuscript}$\}$ $\times$ $\{${Number of submissions}$\}$ $\times$ $\{${Acceptance rate}$\}$.
At this time,
the path coefficients from $\vAnc$ to $\Tre$ were $\coef_{\tre\anc_{1}} = \coef_{\tre\anc_{2}} = 0$
and so the editor considered to intervene these two coefficients $\coef_{\tre\anc_{1}}$ and $\coef_{\tre\anc_{2}}$
to minimize the variance of the number of the pages.
From Section~\ref{subsec:Application-of-Mathematical-Optimization-Procedures-to-Intervention-Effects-for-Variances},
the problem of minimization of the variance can be represented as the following quadratic programming:
\begin{eqnarray}
\Min_{\icoef_{\tre\anc_{1}},\; \icoef_{\tre\anc_{2}}} &
(\icoef_{\tre\anc_{1}}\; \icoef_{\tre\anc_{2}})
\left\{\frac{1}{10} \cdot
\begin{pmatrix}
1 & \frac{1}{10} \\ \frac{1}{10} & 1+\frac{1}{100}
\end{pmatrix}
\right\}
\begin{pmatrix}
\icoef_{\tre\anc_{1}} \\ \icoef_{\tre\anc_{2}}
\end{pmatrix}
+
\left\{\frac{2}{10} \cdot (\frac{1}{10}\;\; \frac{101}{100}) \right\}
\begin{pmatrix}
\icoef_{\tre\anc_{1}} \\ \icoef_{\tre\anc_{2}}
\end{pmatrix}
\nonumber \\
\St & \icoef_{\tre\anc_{1}} \geq -\frac{2}{10},
\nonumber \\
& \icoef_{\tre\anc_{2}} \geq -\frac{2}{10},
\nonumber
\end{eqnarray}
where the constraints for $\icoef_{\tre\anc_{1}}$ and $\icoef_{\tre\anc_{2}}$ were determined
by the editor's inspiration to avoid too strong dependency between $\vAnc$ and $\Tre$.
By computing the above quadratic programming,
the editor obtained the optimal solution $\ocoef_{\tre\anc} = (-0.08, \; -0.20)$
and the variance of $\Obj$ reduced to $0.264$ from $0.301$.
However, the editor noticed that the expectation of the number of pages of the journal
under the optimal solution $\ocoef_{\tre\anc} = (-0.08, \; -0.20)$ is $119.4322$
and thought that it might be too small.
Next,
the editor designated the appropriate amount for the expectation of the number of pages of the journal as $200$
and considered to achieve it by intervention to ${\Mubintre}$.
From Section~\ref{sec:Application-of-Mathematical-Optimization-Procedures-to-Intervention-Effects-for-Variances},
this problem can be formulated as the following quadratic programming:
{\footnotesize
\begin{eqnarray}
\Min_{\iMubintre} &
\iMubintre \cdot 1 \cdot \iMubintre\
+
\left[2 \cdot \left\{ \left( \left( \frac{1}{10} \;\; 1\right) + (\ocoef_{\tre\anc_{1}}\; \ocoef_{\tre\anc_{2}})\begin{pmatrix} 1 & 0 \\ \frac{1}{10} & 1 \end{pmatrix} \right) \begin{pmatrix} \log{10} \\ \log{100} \end{pmatrix}
+  \log{10} - \log{200}\right\}\right]\iMubintre
\nonumber \\
\St & \iMubintre \leq \log{\frac{5}{10}}.
\nonumber
\end{eqnarray}
}
where the constraint for $\iMubintre$ prevents the acceptance rate from exceeding $0.5$.
By computing the above quadratic programming,
the editor obtained the optimal solution $\oMubintre = -0.6931472 = \log{\frac{5}{10}}$.
Then, the expectation of the number of the pages under the optimal solutions
$\ocoef_{\tre\anc} = (-0.08, \; -0.20)$
and $\oMubintre = -0.6931472 = \log{\frac{5}{10}}$
is the $199.0536$.

As a result, the editor succeeded in minimizing the variance of the number of pages of the journal
and adjusting the expectation to the appropriate amount.

What should the editor do, if the editor wants to change the expectation of the number of pages
with the minimized variance?
In this case, all the editor has to do is to re-intervene to ${\Mubintre}$.
The interventions to the path coefficients $\icoef_{\tre\anc_{1}}$ and $\icoef_{\tre\anc_{2}}$ are not needed
because the intervention to ${\Mubintre}$ changes the expectation without changing the minimized variance,
(though, if the constraint for ${\Mubintre}$ is too strong,
then the interventions to $\icoef_{\tre\anc_{1}}$ and $\icoef_{\tre\anc_{2}}$ might be needed to adjust the expectation).
This is the reason why we separate the problem into two algorithms
as in Section~\ref{sec:Application-of-Mathematical-Optimization-Procedures-to-Intervention-Effects-for-Variances}.
Furthermore, note that this two-step procedure has been used in the area of statistical quality control.
\citet{Taguchi1987ExperimentalDesign}
recommended the two-step optimization to solve the design optimization problem,
in which we first maximize the S/N ratio and adjust the expectation on target in the next step.

\section{Conclusion}
\label{sec:Conclusion}

We have introduced matrix representation of total effects and some ideas of their decomposition.
Then, we have shown
that problems
to obtain the optimal intervention that minimizes the variances
and to adjust the expectations
can be formulated as convex quadratic programmings.

In Theorem~\ref{prop:variance_of_descendants},
we assume that $\Cov[\vAnc, \vEpbintre] = \Cov[\vAnc, \vEpbindes] = \Cov[\vTre, \vEpbindes] = \Zero$.
However, this assumption does not hold if there are latent variables
that affect both $\vAnc$ and $\vTre$, or both $\vAnc$ and $\vDes$, or both $\vTre$ and $\vDes$.
In future work,
we intend to extend our results to the case where the assumption of Theorem~\ref{prop:variance_of_descendants} does not hold.

Throughout this paper,
we treat only the case that
the \SEM which represents the true relationships between real objects is given in advance.
Is the method introduced in this paper not useful if we do not have the true model?
We think the answer is yes.
If the given model is not true, then the intervention effect computed by using the method in this paper
and the intervention effect observed in real mostly have different values.
Therefore, the intervention and the computation of the intervention effect based on the given model
can be used for verification whether the model is true or not.
We also intend to consider this subject in future work.

\appendix
\section{Appendix}
\subsection{Kronecker product and Vec Operator}
\label{app:Kronecker_product_and_Vec_Operator}

Let $B = \{b_{ij}\} = [\boldsymbol{b}_{1} \dots \boldsymbol{b}_{n}]$ be an $m \times n$ matrix and
$C$ be a $p \times q$ matrix.

The $mp \times nq$ matrix
\begin{equation}
B \otimes C \define
\begin{pmatrix}
b_{11}C & b_{12}C & \cdots & b_{1n}C \\
b_{21}C & b_{22}C & \cdots & b_{2n}C \\
\vdots  & \vdots  & \ddots & \vdots  \\
b_{m1}C & b_{m2}C & \cdots & b_{mn}C
\end{pmatrix}
\nonumber
\end{equation}
is called the Kronecker product of $B$ and $C$.

The $\veco$ operator for a matrix
is defined as follows.
\begin{equation}
\veco(B) \define
\begin{bmatrix}\boldsymbol{b}_{1} \\ \vdots \\ \boldsymbol{b}_{n} \end{bmatrix}
\nonumber
\end{equation}

Let $D$ be an $n \times p$ matrix.
The following relation holds.
\begin{equation}
\veco(BDC) = (C^{T} \otimes B)\veco(D)
\label{eq:vec_and_otimes}
\end{equation}

\bibliographystyle{optim_sem_arxiv}
\bibliography{optim_sem_arxiv}

\end{document}